\documentclass[aps,prl,preprint,tightenlines,superscriptaddress,showpacs,byrevtex]{revtex4}
\usepackage{epsfig,floatflt}
\usepackage{amssymb,amsmath,citesort,indentfirst,fancyhdr,enumerate}
\usepackage{rotating}

\graphicspath{{ps}}

\def\anti#1{#1{\hspace*{-2.4mm}\rule[3.15mm]{2.3mm}{0.15mm}}\raisebox{-1mm}{}}

\newcommand{\BF}{\ensuremath{{\cal{B}}}}
\newcommand{\Am}{\ensuremath{{\cal{A}}}}
\newcommand{\ACP}{\ensuremath{{A_{CP}}}}
\newcommand{\UFS}{\ensuremath{\Upsilon(4S)}}
\newcommand{\bbbar}{\ensuremath{B\anti{B}}}
\newcommand{\qqbar}{\ensuremath{q\bar{q}}}

\newcommand{\de}{\ensuremath{\Delta E}}
\newcommand{\mb}{\ensuremath{M_{\rm bc}}}

\newcommand{\mass}{MeV/$c^2$}
\newcommand{\Mass}{GeV/$c^2$}
\newcommand{\Masssq}{(GeV/$c^2$)$^2$}

\newcommand{\bckpp}{\ensuremath{B^\pm\to K^\pm\pi^\pm\pi^\mp}}
\newcommand{\bpkpp}{\ensuremath{B^+\to K^+\pi^+\pi^-}}
\newcommand{\bmkpp}{\ensuremath{B^-\to K^-\pi^-\pi^+}}

\newcommand{\Kpp}{\ensuremath{K\pi\pi}}
\newcommand{\kpp}{\ensuremath{K^\pm\pi^\pm\pi^\mp}}
\newcommand{\kppp}{\ensuremath{K^+\pi^+\pi^-}}

\newcommand{\pipi}{\ensuremath{\pi^+\pi^-}}
\newcommand{\kcpi}{\ensuremath{K^+\pi^-}}

\newcommand{\chic}{\ensuremath{\chi_{c0}}}

\newcommand{\sft}{\ensuremath{s_{13}}}
\newcommand{\sst}{\ensuremath{s_{23}}}

\def\nima#1#2#3{{Nucl.\ Instr.\ and Meth.} {\bf A#1}, #3 (#2)}

\def\npb#1#2#3{{ Nucl.\ Phys.}             {\bf B#1}, #3 (#2)}
\def\plb#1#2#3{{ Phys.\ Lett.}             {\bf B#1}, #3 (#2)}

\def\prd#1#2#3{{ Phys.\ Rev.}              {\bf D#1}, #3 (#2)}
\def\pr#1#2#3 {{ Phys.\ Rev.}              {\bf  #1}, #3 (#2)}

\def\prl#1#2#3{{ Phys.\ Rev.\ Lett.}       {\bf  #1}, #3 (#2)}

\begin{document}

\preprint{\vbox{ \hbox{ \vspace*{10mm}  }
                 \hbox{BELLE-CONF-0528}
                 \hbox{LP2005-163}
                 \hbox{EPS05-502} 
}}

\vspace*{20mm}

\title{ \quad\\[0.5cm] 
Search for Direct $CP$ Violation in Three-Body Charmless $\bckpp$ Decay }

\affiliation{Aomori University, Aomori}
\affiliation{Budker Institute of Nuclear Physics, Novosibirsk}
\affiliation{Chiba University, Chiba}
\affiliation{Chonnam National University, Kwangju}
\affiliation{University of Cincinnati, Cincinnati, Ohio 45221}
\affiliation{University of Frankfurt, Frankfurt}
\affiliation{Gyeongsang National University, Chinju}
\affiliation{University of Hawaii, Honolulu, Hawaii 96822}
\affiliation{High Energy Accelerator Research Organization (KEK), Tsukuba}
\affiliation{Hiroshima Institute of Technology, Hiroshima}
\affiliation{Institute of High Energy Physics, Chinese Academy of Sciences, Beijing}
\affiliation{Institute of High Energy Physics, Vienna}
\affiliation{Institute for Theoretical and Experimental Physics, Moscow}
\affiliation{J. Stefan Institute, Ljubljana}
\affiliation{Kanagawa University, Yokohama}
\affiliation{Korea University, Seoul}
\affiliation{Kyoto University, Kyoto}
\affiliation{Kyungpook National University, Taegu}
\affiliation{Swiss Federal Institute of Technology of Lausanne, EPFL, Lausanne}
\affiliation{University of Ljubljana, Ljubljana}
\affiliation{University of Maribor, Maribor}
\affiliation{University of Melbourne, Victoria}
\affiliation{Nagoya University, Nagoya}
\affiliation{Nara Women's University, Nara}
\affiliation{National Central University, Chung-li}
\affiliation{National Kaohsiung Normal University, Kaohsiung}
\affiliation{National United University, Miao Li}
\affiliation{Department of Physics, National Taiwan University, Taipei}
\affiliation{H. Niewodniczanski Institute of Nuclear Physics, Krakow}
\affiliation{Nippon Dental University, Niigata}
\affiliation{Niigata University, Niigata}
\affiliation{Nova Gorica Polytechnic, Nova Gorica}
\affiliation{Osaka City University, Osaka}
\affiliation{Osaka University, Osaka}
\affiliation{Panjab University, Chandigarh}
\affiliation{Peking University, Beijing}
\affiliation{Princeton University, Princeton, New Jersey 08544}
\affiliation{RIKEN BNL Research Center, Upton, New York 11973}
\affiliation{Saga University, Saga}
\affiliation{University of Science and Technology of China, Hefei}
\affiliation{Seoul National University, Seoul}
\affiliation{Shinshu University, Nagano}
\affiliation{Sungkyunkwan University, Suwon}
\affiliation{University of Sydney, Sydney NSW}
\affiliation{Tata Institute of Fundamental Research, Bombay}
\affiliation{Toho University, Funabashi}
\affiliation{Tohoku Gakuin University, Tagajo}
\affiliation{Tohoku University, Sendai}
\affiliation{Department of Physics, University of Tokyo, Tokyo}
\affiliation{Tokyo Institute of Technology, Tokyo}
\affiliation{Tokyo Metropolitan University, Tokyo}
\affiliation{Tokyo University of Agriculture and Technology, Tokyo}
\affiliation{Toyama National College of Maritime Technology, Toyama}
\affiliation{University of Tsukuba, Tsukuba}
\affiliation{Utkal University, Bhubaneswer}
\affiliation{Virginia Polytechnic Institute and State University, Blacksburg, Virginia 24061}
\affiliation{Yonsei University, Seoul}
  \author{K.~Abe}\affiliation{High Energy Accelerator Research Organization (KEK), Tsukuba} 
  \author{K.~Abe}\affiliation{Tohoku Gakuin University, Tagajo} 
  \author{I.~Adachi}\affiliation{High Energy Accelerator Research Organization (KEK), Tsukuba} 
  \author{H.~Aihara}\affiliation{Department of Physics, University of Tokyo, Tokyo} 
  \author{K.~Aoki}\affiliation{Nagoya University, Nagoya} 
  \author{K.~Arinstein}\affiliation{Budker Institute of Nuclear Physics, Novosibirsk} 
  \author{Y.~Asano}\affiliation{University of Tsukuba, Tsukuba} 
  \author{T.~Aso}\affiliation{Toyama National College of Maritime Technology, Toyama} 
  \author{V.~Aulchenko}\affiliation{Budker Institute of Nuclear Physics, Novosibirsk} 
  \author{T.~Aushev}\affiliation{Institute for Theoretical and Experimental Physics, Moscow} 
% ...
\collaboration{The Belle Collaboration}

\begin{abstract}
We report results on studies of $CP$ violation in the three-body charmless
decay $\bckpp$. Evidence at the $3.9\sigma$ level for large direct $CP$
violation in $B^\pm\to\rho(770)^0K^\pm$ is found. This is the first evidence
for $CP$  violation in a charged meson decay. The analysis is performed using
Dalitz analysis technique with a data sample that contains 386 million
$B\bar{B}$ pairs collected near the $\Upsilon(4S)$ resonance, with the Belle
detector operating at the KEKB asymmetric energy $e^+ e^-$ collider.
\end{abstract}


\maketitle

{\renewcommand{\thefootnote}{\fnsymbol{footnote}}}
\setcounter{footnote}{0}


\vspace*{60mm}

\section{Introduction}

Decays of $B$ mesons to three-body charmless hadronic final states may provide
new possibilities for $CP$ violation searches. In contrast to decays to
two-body final states where direct $CP$ violation can only manifest itself as
difference in decay rates for $B$ and $\bar{B}$ mesons to charge
conjugate final states, in three-body decays it can also be observed as a
difference in relative phases between two quasi-two-body channels.
A necessary condition for observing direct $CP$ violation
in a two-body decay is a non-trivial strong phase difference between
the $CP$-conserving and $CP$-violating amplitudes contributing to a particular
final state. Although this condition (if satisfied) also enhances the
sensitivity to $CP$ violation in three-body decays, it is not required
in general and direct $CP$ violation in quasi-two-body decays can also be
observed with any strong phase difference via the interference with a nearby
quasi-two-body or non-resonant amplitude(s). Although direct $CP$ violation
has been observed in decays of neutral $K$ mesons~\cite{dcpv-K0} and recently
in neutral $B$ meson decays~\cite{dcpv-B0} no $CP$ violation in decays of
charged mesons has been found to date. However, large direct $CP$ violation
is expected in some quasi-two-body modes~\cite{beneke-neubert}. Several other
ideas to study $CP$ violation utilizing decays to three-body final states
have been proposed~\cite{gronau,b2hhhcp,garmash2,hazumi}.

First results on the amplitude analysis of the $\bckpp$ decay are described in 
Refs.~\cite{khh-dalitz-belle,hhh-dalitz-babar}; the first results on searches
for direct $CP$ violation from independent fits of the $B^-$ and $B^+$ samples
are given in Ref.~\cite{belle-kpp-dcpv,babar-kpp-dcpv}. The analysis of direct
$CP$ violation in the decay $\bckpp$ described in this paper is based on a
simultaneous fit to $B^-$ and $B^+$ events. The results are obtained with a
data sample of 357\,fb$^{-1}$ containing 386 million $B\bar{B}$ pairs,
collected with the Belle detector operating at the KEKB asymmetric-energy
$e^+e^-$ collider~\cite{KEKB} with a center-of-mass (c.m.)\
energy at the $\Upsilon(4S)$ resonance (on-resonance data). The beam energies
are 3.5 GeV for positrons and 8.0 GeV for electrons. For the study of the
$e^+e^-\to q\bar{q}$ continuum background, we use data taken 60~MeV below the
$\Upsilon(4S)$ resonance (off-resonance data).


\section{The Belle detector}

  The Belle detector~\cite{Belle} is a large-solid-angle magnetic spectrometer
based on a 1.5~T superconducting solenoid magnet. Charged particle tracking is
provided by a silicon vertex detector and a 50-layer central drift chamber
(CDC) that surround the interaction point. Two inner detector configurations
were used. A 2.0 cm beampipe and a 3-layer silicon vertex detector was used
for the first sample of 152 million $B\bar{B}$ pairs, while a 1.5 cm beampipe,
a 4-layer silicon detector and a small-cell inner drift chamber were used to
record the remaining 234 million $B\bar{B}$ pairs~\cite{Ushiroda}. The charged
particle acceptance covers laboratory polar angles between $\theta=17^{\circ}$
and $150^{\circ}$, corresponding to about 92\% of the total solid angle in
the c.m.\ frame. The momentum resolution is determined from cosmic rays and
$e^+ e^-\to\mu^+\mu^-$ events to be $\sigma_{p_t}/p_t=(0.30\oplus0.19p_t)\%$,
where $p_t$ is the transverse momentum in GeV/$c$.

  Charged hadron identification is provided by $dE/dx$ measurements in the CDC,
an array of 1188 aerogel \v{C}erenkov counters (ACC), and a barrel-like array
of 128 time-of-flight scintillation counters (TOF); information from the three
subdetectors is combined to form a single likelihood ratio, which is then used
in kaon and pion selection. At large momenta (\mbox{$>2.5$}~GeV/$c$) only the
ACC and CDC are used to separate charged pions and kaons since here the TOF
provides no additional discrimination. Electromagnetic showering particles are
detected in an array of 8736 CsI(Tl) crystals (ECL) that covers the same solid
angle as the charged particle tracking system. The energy resolution for
electromagnetic showers is
$\sigma_E/E = (1.3 \oplus 0.07/E \oplus 0.8/E^{1/4})\%$, where $E$ is in GeV.
Electron identification in Belle is based on a combination of $dE/dx$
measurements in the CDC, the response of the ACC, and the position, shape and
total energy deposition (i.e., $E/p$) of the shower detected in the ECL.
The electron identification efficiency is greater than 92\% for tracks with
$p_{\rm lab}>1.0$~GeV/$c$ and the hadron misidentification probability is below
0.3\%. The magnetic field is returned via an iron yoke that is instrumented to
detect muons and $K^0_L$ mesons. We use a GEANT-based Monte Carlo (MC)
simulation to model the response of the detector and determine its
acceptance~\cite{GEANT}.


\section{Event Reconstruction}

Charged tracks are selected with a set of track quality requirements based
on the number of CDC hits and on the distances of closest approach to the
interaction point. We also require that the track momenta transverse to the
beam be greater than 0.1~GeV/$c$ to reduce the low momentum combinatorial
background. For charged kaon identification we impose a requirement on the
particle identification variable which has 86\% efficiency and a 7\% fake
rate from misidentified pions. Charged tracks that are positively identified
as electrons or protons are excluded. Since the muon identification efficiency
and fake rate vary significantly with the track momentum, we do not veto muons
to avoid additional systematic errors.

We identify $B$ candidates using two variables: the energy difference
$
\de = E_B - E^*_{\rm beam} = 
(\sum_i\sqrt{c^2{\bf p}_i^2 + c^4m_i^2} ) - E^*_{\rm beam},
$
and the beam constrained mass
$
\mb =  \frac{1}{c^2}\sqrt{E^{*2}_{\rm beam}-c^2(\sum_i {\bf p}_i)^2},
$
where the summation is over all particles from a $B$ candidate; ${\bf p}_i$
and $m_i$ are their c.m.\ three-momenta and masses, respectively.
The signal $\de$ shape is fitted by a sum of two Gaussian functions with a
common mean. In fits to the experimental data, we fix the width and the
relative fraction of the second Gaussian function from MC simulation. The
common mean of the two Gaussian functions and the width of the main Gaussian
are floated. The $\de$ shape for the $\qqbar$ background is parametrized by
a linear function.
The $\mb$ distribution for the signal events is parametrized by a single
Gaussian function. The $\mb$ width is about 3~MeV/$c^2$ and, in general, does
not depend on the final state (unless photons are included in the reconstructed
final state). The background shape is parametrized with an empirical function
$f(\mb)\propto x\sqrt{1-x^2}\exp[-\xi(1-x^2)]$~\cite{ArgusF}, where
$x = \mb/E^*_{\rm beam}$ and $\xi$ is a parameter.


\section{Background Suppression}

The dominant background is due to $e^+e^-\to~\qqbar$ ($q = u, d, s$ and $c$
quarks) continuum events that have a cross-section about three times larger
than that for the $e^+e^-\to\UFS\to\bbbar$. This background is suppressed
using variables that characterize the event topology. Since the two $B$
mesons produced from an $\UFS$ decay are nearly at rest in the c.m.\ frame,
their decay products are uncorrelated and the event tends to be spherical. In
contrast, hadrons from continuum $\qqbar$ events tend to exhibit a two-jet
structure. We use $\theta_{\rm thr}$, which is the angle between the thrust
axis of the $B$ candidate and that of the rest of the event, to discriminate
between the two cases. The distribution of $|\cos\theta_{\rm thr}|$ is
strongly peaked near $|\cos\theta_{\rm thr}|=1.0$ for $\qqbar$ events and is
nearly flat for $\bbbar$ events. A Fisher discriminant is utilized for the
further suppression of the continuum background. When combined, these two
variables reject about 98\% of the continuum background in the $\bckpp$ decay
while retaining 36\% of the signal. A detailed description of the continuum
suppression technique can be found in Ref.~\cite{garmash2} and references
therein.

Another background originates from other $B$ meson decays. We study the
$\bbbar$ related background using a large sample of MC generated $\bbbar$
generic events. Note that charmless hadronic $B$ decays that proceed
via $b\to s(d)$ penguin and $b\to u$ tree transitions are not included in the
generic MC sample and are generated separately. We find that the dominant
$\bbbar$ related background to the $\kppp$ final state is due to
$B^+\to \anti{D}^0\pi^+$, $\anti{D}^0\to K^+\pi^-$ and due to
$B^+\to J/\psi(\psi(2S))K^+$, $J/\psi(\psi(2S))\to \mu^+\mu^-$ decays. We veto
$B^+\to \anti{D}^0\pi^+$ events by requiring $|M(K\pi)-M_D|>100$~MeV/$c^2$.
We also veto $B^+\to \anti{D}^0K^+$, $\anti{D}^0\to\pipi$ signal by
requiring $|M(\pipi)-M_D|>15$~MeV/$c^2$.
To suppress the background due to $\pi/K$ misidentification, we also exclude
candidates if the invariant mass of any pair of oppositely charged tracks from
the $B$ candidate is consistent with the $\bar{D}^0\to K^+\pi^-$ hypothesis
within 25~MeV/$c^2$ ($\sim 4\sigma$), regardless of the particle
identification information. Modes with $J/\psi(\psi(2S))$ in the final state
contribute due to muon-pion misidentification; the contribution from the
$J/\psi(\psi(2S))\to e^+e^-$ submode is found to be negligible after the
electron veto requirement. We exclude $J/\psi(\psi(2S))$ background by
requiring $|M(\pi^+\pi^-)_{\mu^+\mu^-}-M_{J/\psi}|>70$~MeV/$c^2$ and
$|M(\pi^+\pi^-)_{\mu^+\mu^-}-M_{\psi(2S)}|>50$~MeV/$c^2$, with a muon mass
assignment used here for the pion candidates. 
Yet another small but clearly visible background is due to $B^+\to J/\psi K^+$,
$J/\psi\to \mu^+\mu^-$ decay with a complicated series of particle
misidentifications; the charged kaon from the $B$ is misidentified as a pion,
the $\mu^+$ is misidentified as a kaon and the $\mu^-$ as another pion. This
background is excluded by applying a veto on the invariant mass of oppositely
charged kaon and pion candidates:
$|M(K^+\pi^-)_{\mu^+\mu^-}-M_{J/\psi}|>20$~MeV/$c^2$. The most significant
background from charmless $B$ decays is found to originate from
$B^+\to\eta'K^+$ followed by $\eta'\to\pi^+\pi^-\gamma$. Another contribution
comes from $B^+\to\pi^+\pi^+\pi^-$ decay, where one of the two same charge
pions is misidentified as a kaon. Finally, we consider a background from the
decay $B^0\to K^+\pi^-$. Although it does not directly contribute to the $\de$
signal region, this background should be taken into account in order to
correctly estimate the $\qqbar$ component of the background.


\begin{figure}[!t]
 \includegraphics[width=0.32\textwidth]{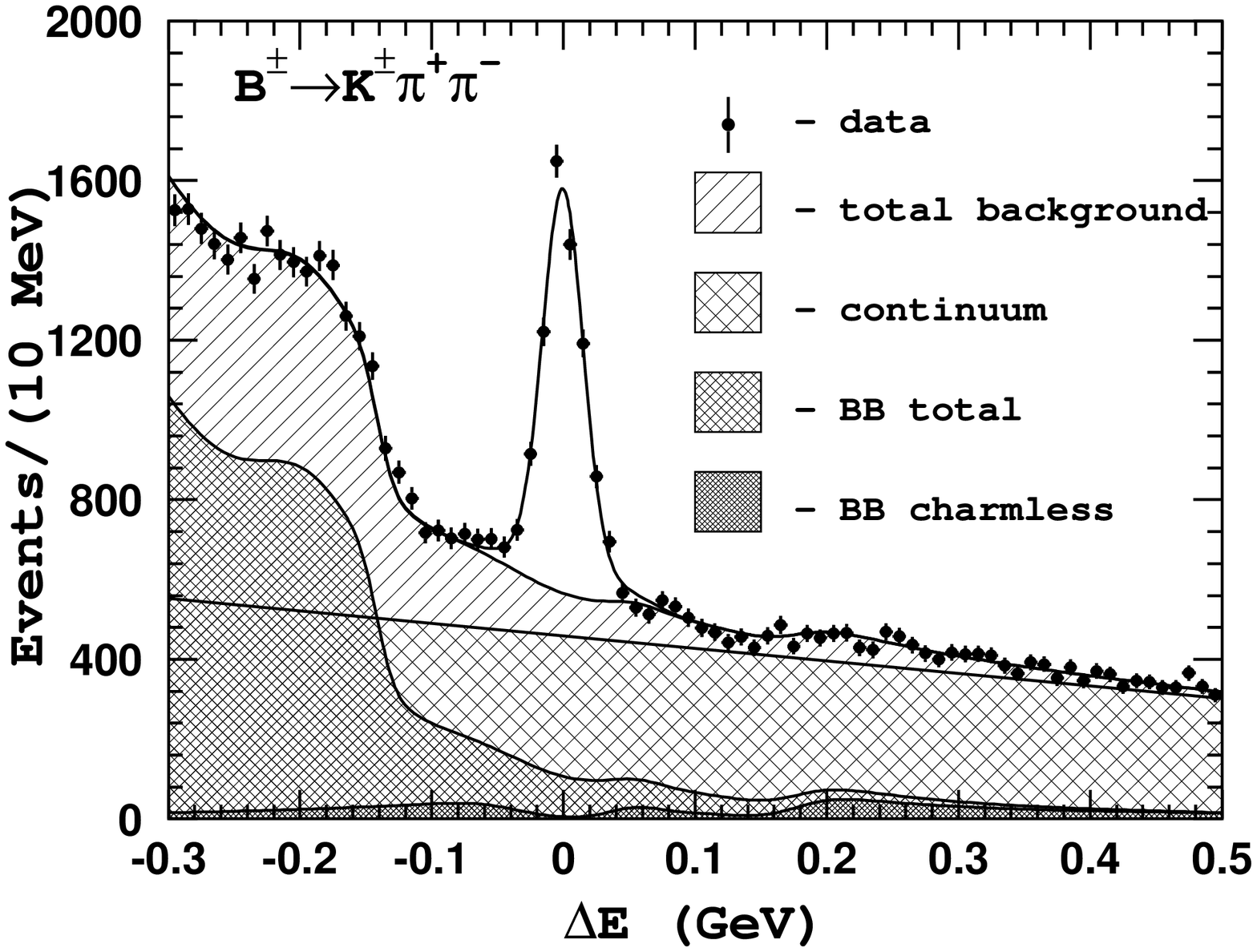} \hfill
 \includegraphics[width=0.32\textwidth]{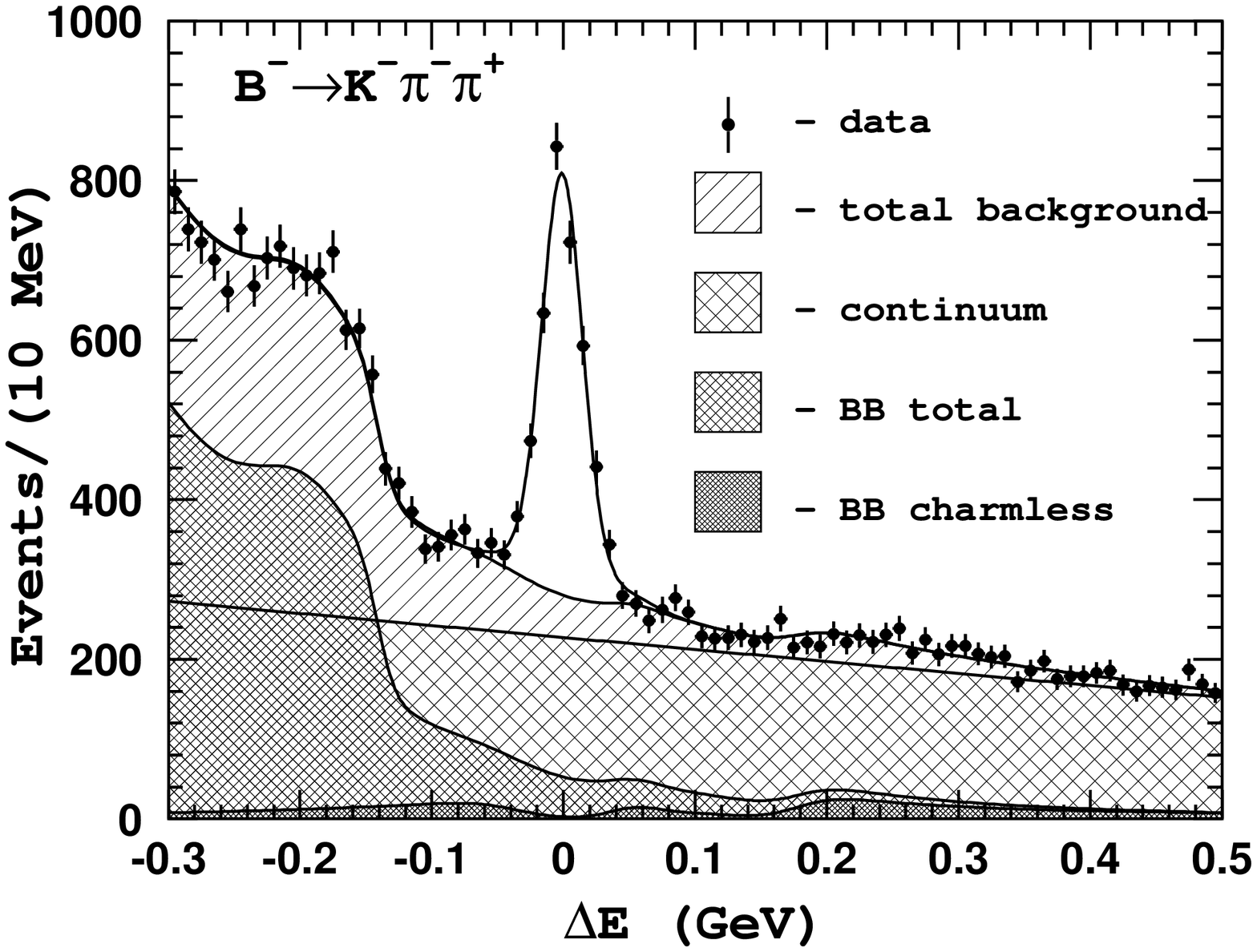} \hfill
 \includegraphics[width=0.32\textwidth]{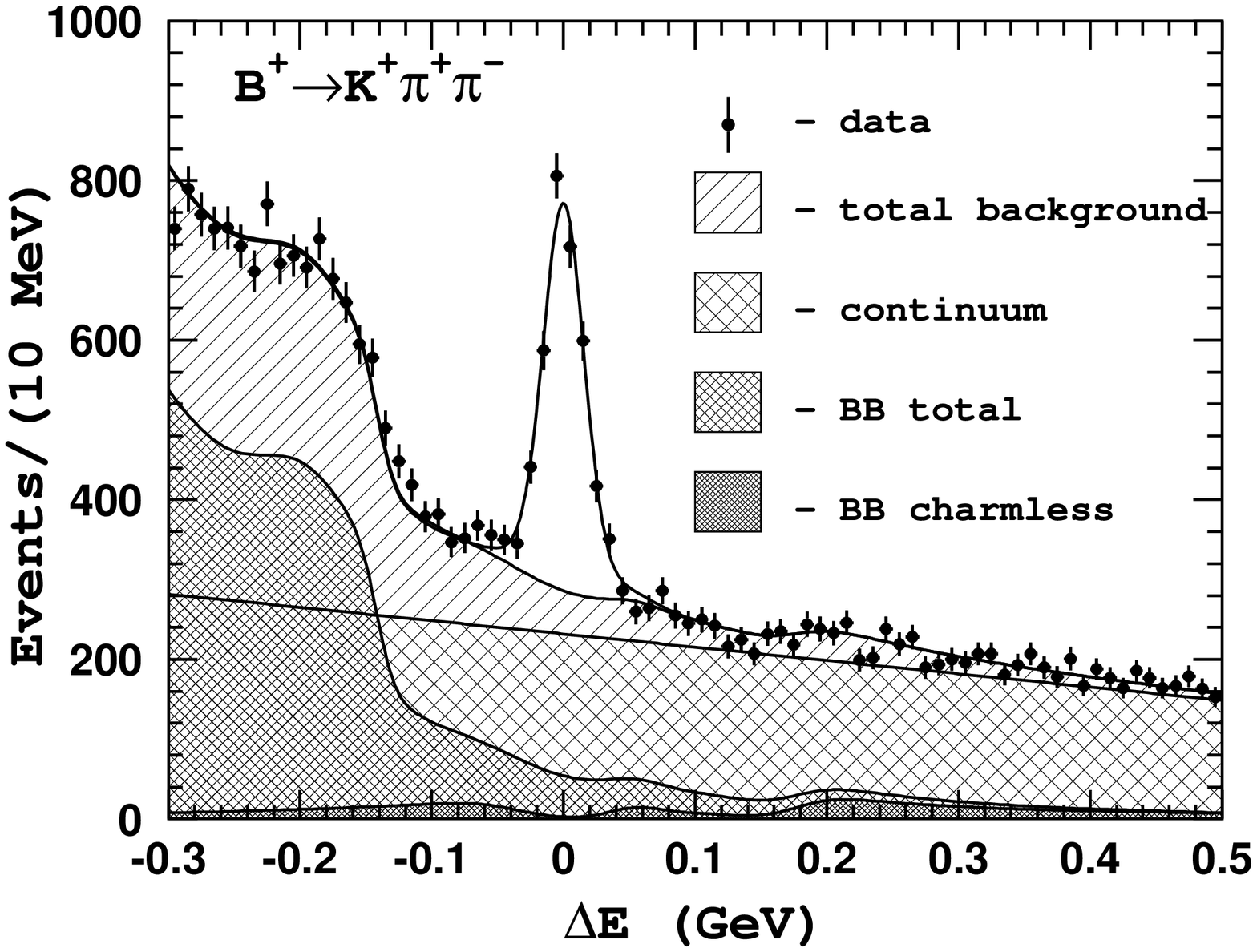}
 \caption{$\de$ distributions for (a) $\bckpp$ combined; (b) $\bmkpp$ and
          (c) $\bpkpp$ candidate events.
          Points with error bars are data; the open histogram is the fit 
          result; the hatched histograms are various background components.}
  \label{fig:khh-DE}
\end{figure}

\section{Three-body Signal Yields}
\label{sec:khh}

The $\de$ distribution for $\bckpp$ candidates that pass all the selection
requirements are shown in Fig.~\ref{fig:khh-DE}. In the fit to the $\de$
distribution we fix the shape and
normalization of the charmless $\bbbar$ background components from the measured
branching fractions~\cite{PDG} and known number of produced $\bbbar$ events.
For the $\bbbar$ generic component we fix only the shape and let the
normalization float. The slope and normalization of the $\qqbar$ background
component are free parameters. Results of the fit are shown in
Fig.~\ref{fig:khh-DE}, where different components of the background are shown
separately for comparison. There is a large increase in the level of $\bbbar$
related background in the region $\de<-0.15$~GeV. This is mainly due to
$B\to D\pi$, $D\to K\pi\pi$ decay. This decay mode produces the same final
state as the studied process plus one extra pion that is not included in the
energy difference calculation. The semileptonic decays $B\to D^{(*)}\pi$,
$D\to K\mu\nu_\mu$ also contribute due to muon-pion misidentification. The
shape of these backgrounds is well described by MC simulation. Results of the
$\de$ fits are given in Table~\ref{tab:defitall}.

\begin{table}[!b]
  \caption{Results of the fits of the $\de$ distributions.}
  \medskip
  \label{tab:defitall}
\centering
  \begin{tabular}{lccccccc} \hline \hline
\hspace*{0mm}  Final state       \hspace*{1mm} & 
\hspace*{2mm}  $\sigma_1$        \hspace*{2mm} &
\hspace*{2mm}  $\sigma_2$        \hspace*{2mm} &
\hspace*{2mm}  Fraction of the   \hspace*{2mm} &
\hspace*{2mm}  $N_{\qqbar}$      \hspace*{2mm} &
\hspace*{2mm}  $N_{B\bar{B}}$    \hspace*{2mm} &
\hspace*{2mm}  Signal Yield      \hspace*{1mm} \\
     &  MeV & MeV & main Gaussian & (events)  & (events)  & (events)    \\
\hline
$~\bmkpp$ & $15.6\pm0.6$ & $ 35.0$ (fixed) & $0.84$ (fixed) & $16932\pm275$ &  $8639\pm226$ & $2248\pm79$ \\
$~\bpkpp$ & $15.0\pm0.6$ & $ 35.0$ (fixed) & $0.84$ (fixed) & $17268\pm274$ &  $8828\pm227$ & $2038\pm76$ \\
\hline
$~\bckpp$ & $15.3\pm0.5$ & $ 35.0$ (fixed) & $0.84$ (fixed) & $34188\pm386$ & $17452\pm320$ & $4286\pm99$ \\
\hline \hline
  \end{tabular}
\end{table}

\begin{figure}[!t]
 \includegraphics[width=0.47\textwidth]{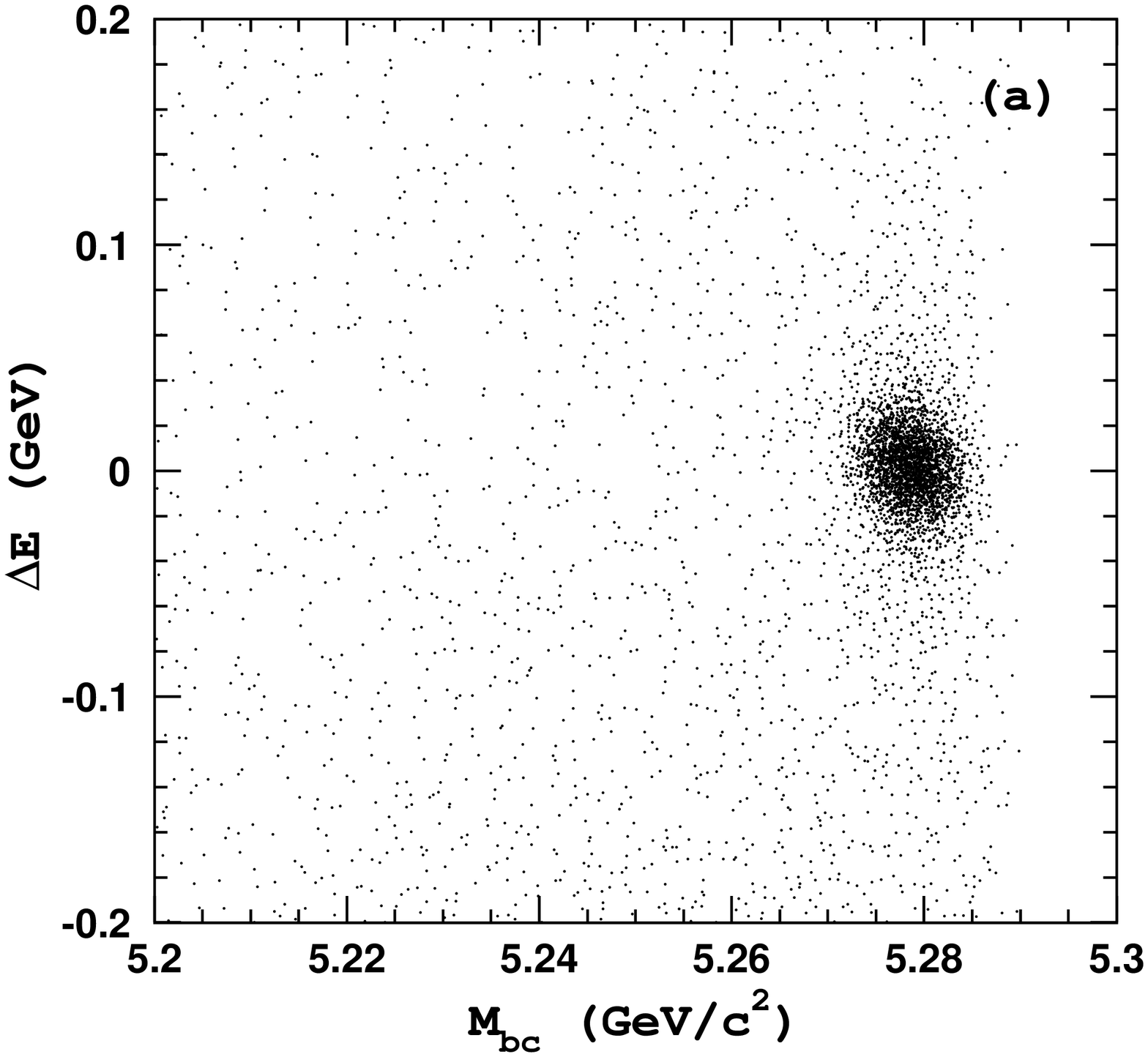} \hfill
 \includegraphics[width=0.47\textwidth]{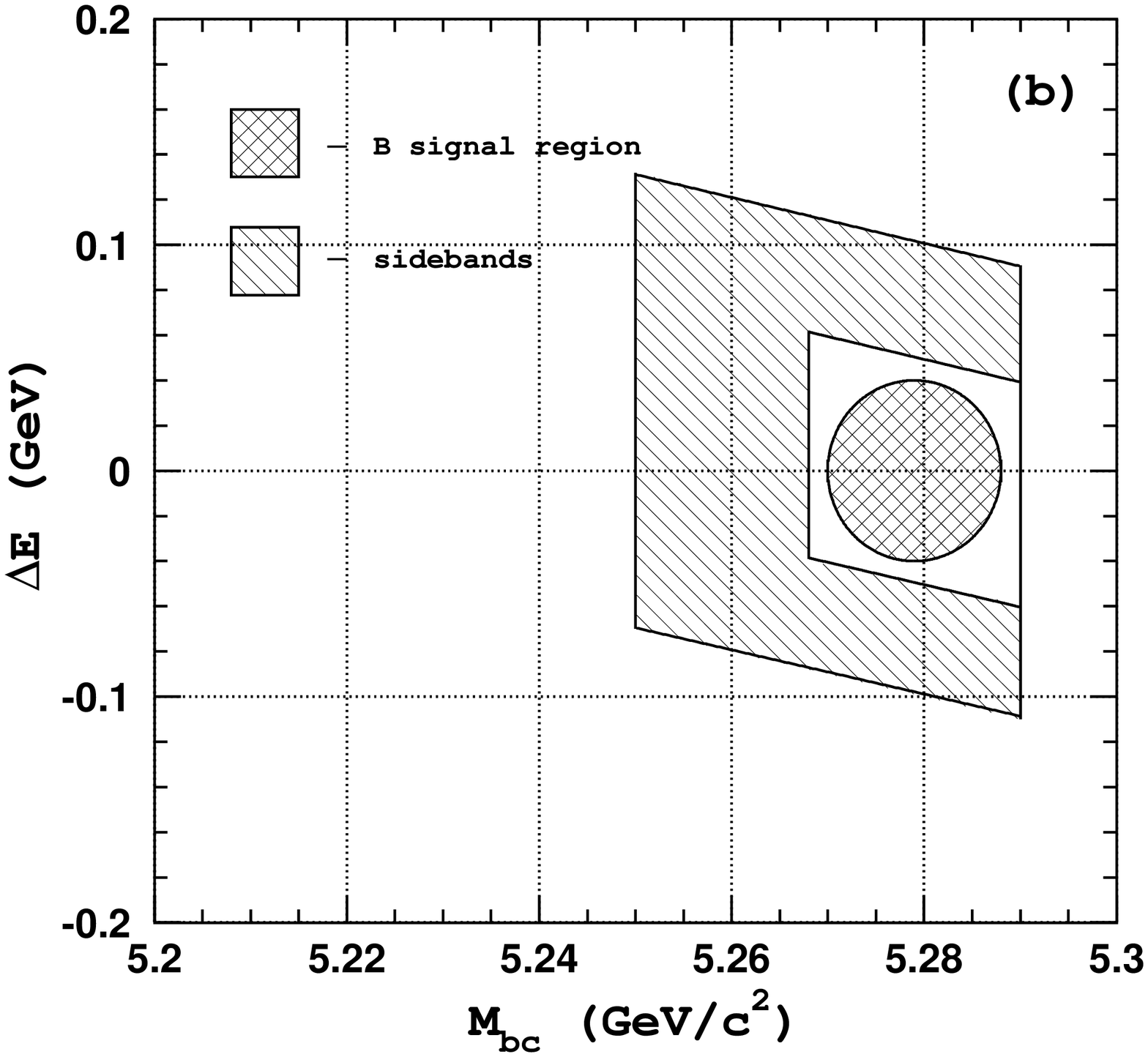}
 \caption{(a) Two-dimensional $\de$ versus $\mb$ plot for the $\bckpp$
          signal MC events; \mbox{(b) Definition} of the signal and
          sideband regions on the $\mb-\de$ plane.}
 \label{fig:dE-Mbc}
\end{figure}

For the analysis of  quasi-two-body intermediate states that contribute to
the observed $\bckpp$ three-body signal, we define the $B$ signal and
sideband regions as shown in Fig.~\ref{fig:dE-Mbc}. Defined in this way, the
$\mb-\de$ sidebands are equivalent to the following sidebands in terms of the
three-particle invariant mass $M(K\pi\pi)$ and three-particle momentum
$P(K\pi\pi)$:
\[
0.05 {\rm ~GeV}/c^2 < |M(K\pi\pi)-M_B| < 0.10 {\rm ~GeV}/c^2;
~~~P(K\pi\pi) < 0.48 {\rm ~GeV}/c.~~~~~~~  \nonumber
\]
and
\[
|M(K\pi\pi)-M_B| < 0.10 {\rm ~GeV}/c^2;
~~~0.48 {\rm ~GeV}/c < P(K\pi\pi) < 0.65 {\rm ~GeV}/c.~~~~~~~  \nonumber
\]

  The signal region is defined as an ellipse around the $\mb$ and $\de$ mean
values:
\[
\frac{(\mb-M_B)^2}{(7.5~{\rm MeV}/c^2)^2} + \frac{\de^2}{(40~{\rm MeV})^2} < 1,
\]
The efficiency of the requirements that define the signal region is 0.927;
the total number of events in the signal region is 7757. The relative fraction
of signal events in the $B$ signal region is determined to be $0.512\pm0.012$.
There are 27855 events in the sideband region that is about seven times the
estimated number of background events in the signal region.


\section{Amplitude Analysis}

The amplitude analysis of three-body $B$ meson decay reported here is
performed by means of an unbinned maximum likelihood fit. Details of the
analysis technique are described in Ref.~\cite{khh-dalitz-belle}. One of the
important questions
that arise in unbinned analysis is the estimation of the goodness-of-fit.
As the unbinned maximum likelihood fitting method does not provide a direct
way to estimate the quality of the fit, we need a measure to assess how well
any given fit represents the data. To do so the following procedure is
applied. We first subdivide the entire Dalitz plot into
1~\Masssq$\times$1~\Masssq
bins. If the number of events in the bin is smaller than $N_{\rm min}=25$
it is combined with the adjacent bins until the number of events exceed the
minimum required level. After completing this procedure, the entire Dalitz
plot is divided into set of bins of varying size, and a $\chi^2$ variable
for the multinomial distribution can be calculated as
\begin{equation}
   \chi^2 = -2\sum^{N_{\rm bins}}_{i=1}n_i\ln\left(\frac{p_i}{n_i}\right),
\end{equation}
where $n_i$ is the number of events observed in $i$-th bin, and $p_i$ is the
number of predicted events from the fit. For a large number of events this
formulation becomes equivalent to the usual one. Since we are minimizing the
unbinned likelihood function, our ``$\chi^2$'' variable does not
asymptotically follow a $\chi^2$ distribution but it is bounded by a $\chi^2$
variable with ($N_{\rm bins}-1$) degrees of freedom and a $\chi^2$ variable
with ($N_{\rm bins}-k-1$) degrees of freedom, where $k$ is the number of fit
parameters. Because it is bounded by two $\chi^2$ variables, it should be a
useful statistic for comparing the relative goodness of fits for different
models.

\begin{figure}[!t]
  \centering
  \includegraphics[width=0.48\textwidth]{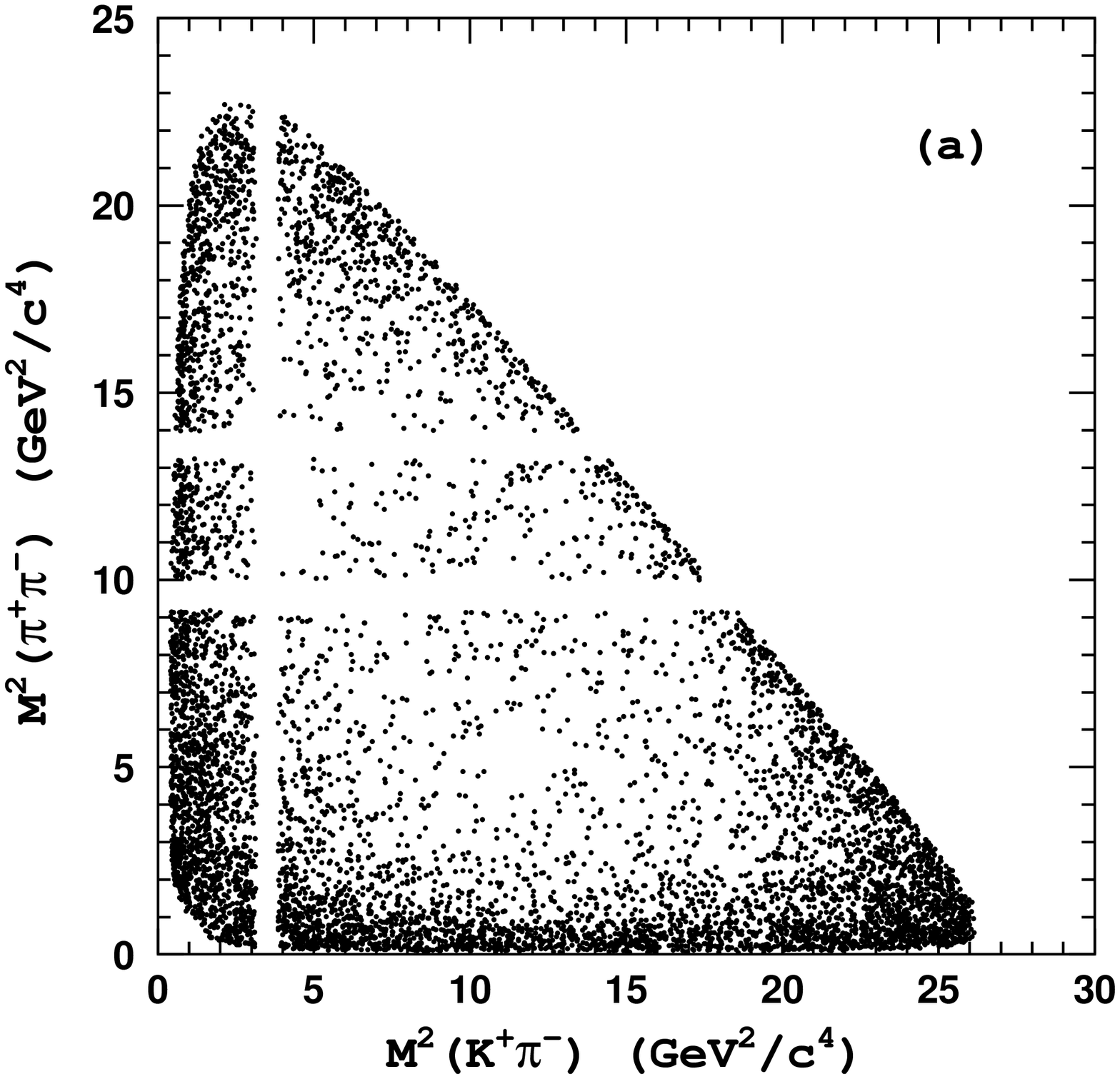} \hfill
  \includegraphics[width=0.48\textwidth]{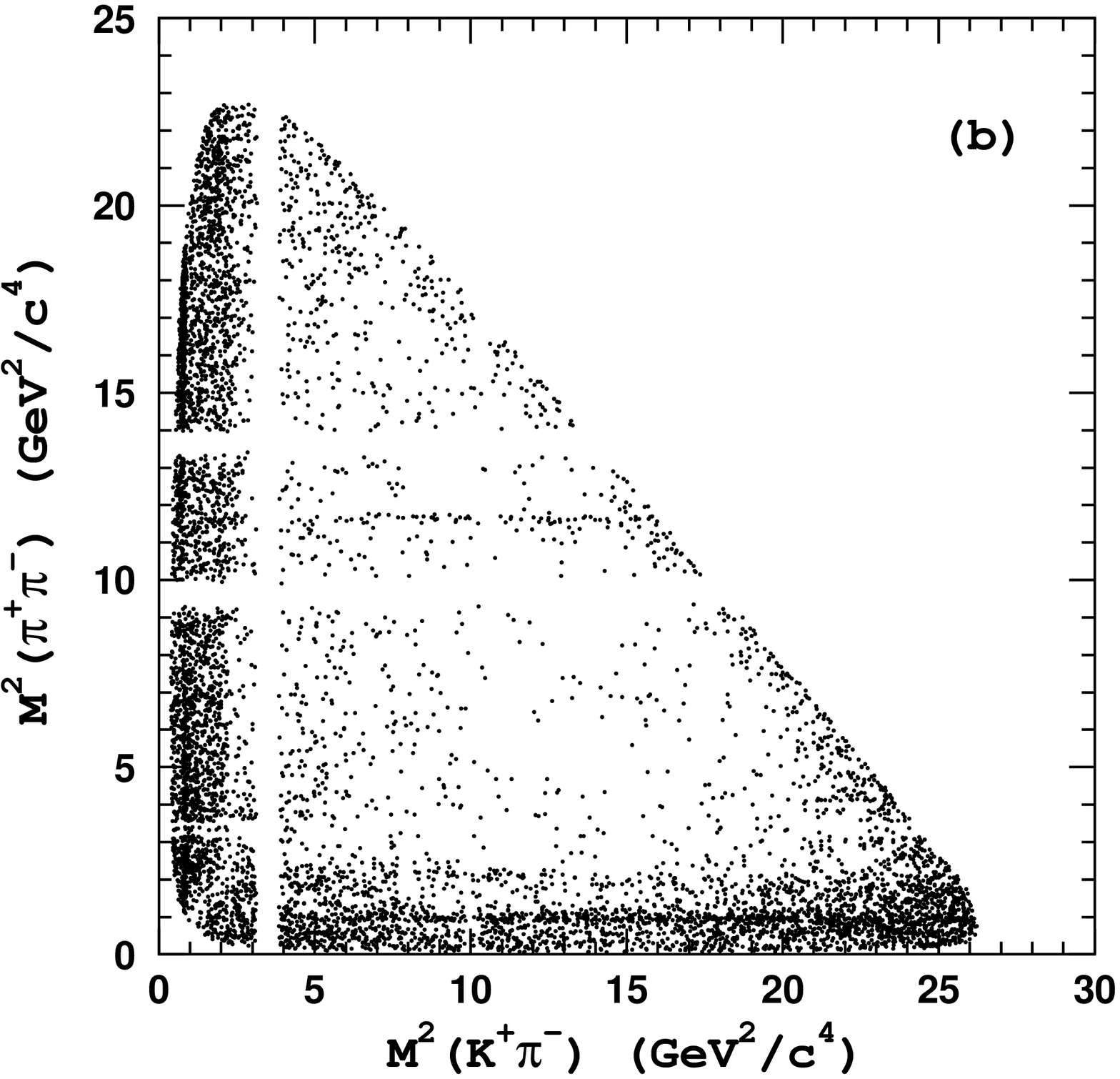}
  \caption{Dalitz plot for the $\kpp$ events in the (a) $\mb-\de$ sideband
           region and in (b) $B$ signal region.}
\label{fig:kpp_dp}
\end{figure}


\subsection{Fitting the Background Shape}
\label{sec:khh-bac}

  Before fitting the Dalitz plot for events in the signal region, we need to
determine the distribution of background events. The background density
function is determined from an unbinned likelihood fit to the events in the
$\mb-\de$ sidebands defined in Fig.~\ref{fig:dE-Mbc}.
Figure~\ref{fig:kpp_dp}(a) shows the Dalitz plot for sideband events.

We use the following empirical parameterization to describe the distribution
of background events over the Dalitz plot
\begin{eqnarray}
B_{\Kpp}(\sft,\sst) &=& \alpha_1e^{-\beta_1s_{13}}
          ~+~ \alpha_2e^{-\beta_2s_{23}}
          ~+~ \alpha_3e^{-\beta_3s_{12}} \nonumber \\
          &+& \alpha_4e^{-(\beta_4s_{13}+\beta_5s_{23})}
          ~+~ \alpha_5e^{-(\beta_6s_{13}+\beta_7s_{12})}
          ~+~ \alpha_6e^{-(\beta_8s_{23}+\beta_9s_{12})} \nonumber \\
          &+& \gamma_1 |BW(K^*(892))|^2 ~+~ \gamma_2 |BW(\rho(770))|^2,
\label{eq:kpp_back}
\end{eqnarray}
where $s_{13} \equiv M^2(\kcpi)$, $s_{23}\equiv M^2(\pipi)$ and $\alpha_i$
($\alpha_1\equiv 1.0$), $\beta_i$ and $\gamma_i$ are fit parameters; $BW$
is a Breit-Wigner function. 
The first three terms in Eq.~(\ref{eq:kpp_back}) are introduced to describe
the background enhancement in the two-particle low invariant mass regions.
This enhancement originates mainly from $e^+e^-\to\qqbar$ continuum
events. Due to the jet-like structure of this background, all three particles
in a three-body combination have almost collinear momenta. Hence, the
invariant mass of at least one pair of particles is in the low mass region.
In addition, it is often the case that two high momentum particles are
combined with a low momentum particle to form a $B$ candidate. In this
case there are two pairs with low invariant masses and one pair with high
invariant mass. This results in even  stronger enhancement of the background
in the corners of the Dalitz plot. This is taken into account by terms $4-6$
in Eq.~(\ref{eq:kpp_back}). To account for the contribution from real
$K^*(892)^0$ and $\rho(770)^0$ mesons, we introduce two more terms in
Eq.~(\ref{eq:kpp_back}), that are (non-interfering) squared Breit-Wigner
amplitudes, with masses and widths fixed at world average values~\cite{PDG}.

\begin{figure}[!t]
  \centering
  \includegraphics[width=0.33\textwidth]{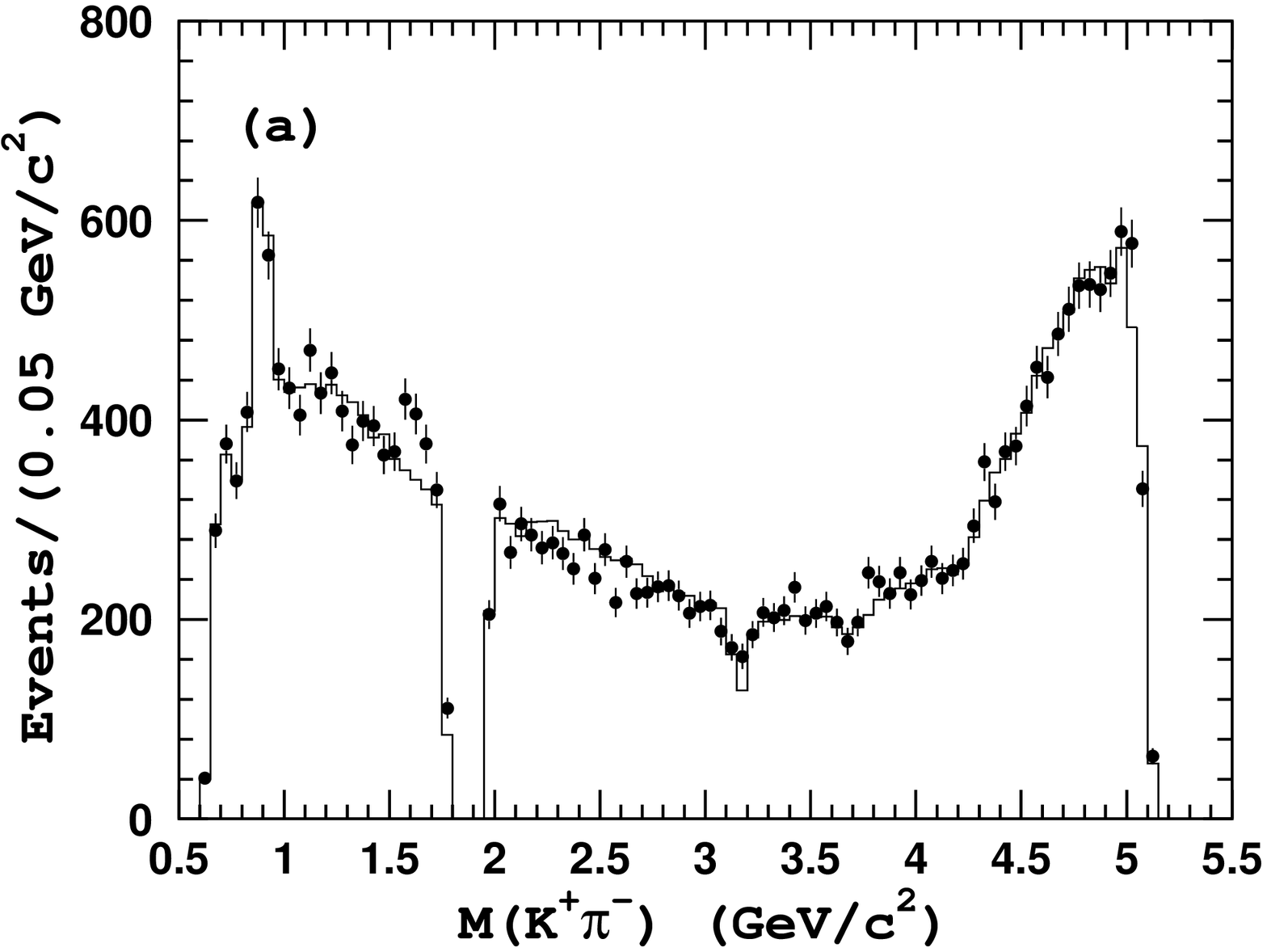} \hspace*{-2mm}
  \includegraphics[width=0.33\textwidth]{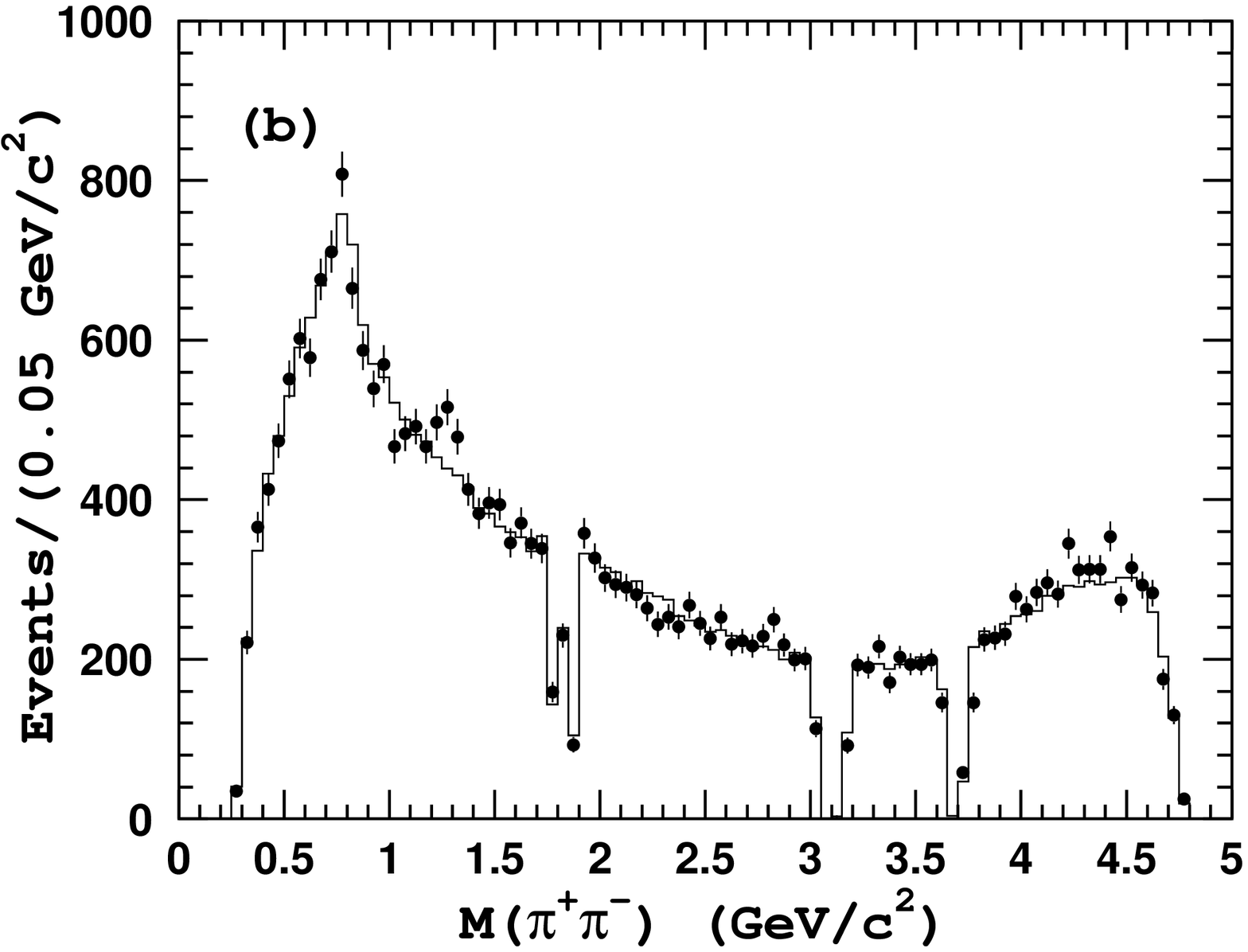} \hspace*{-2mm}
  \includegraphics[width=0.33\textwidth]{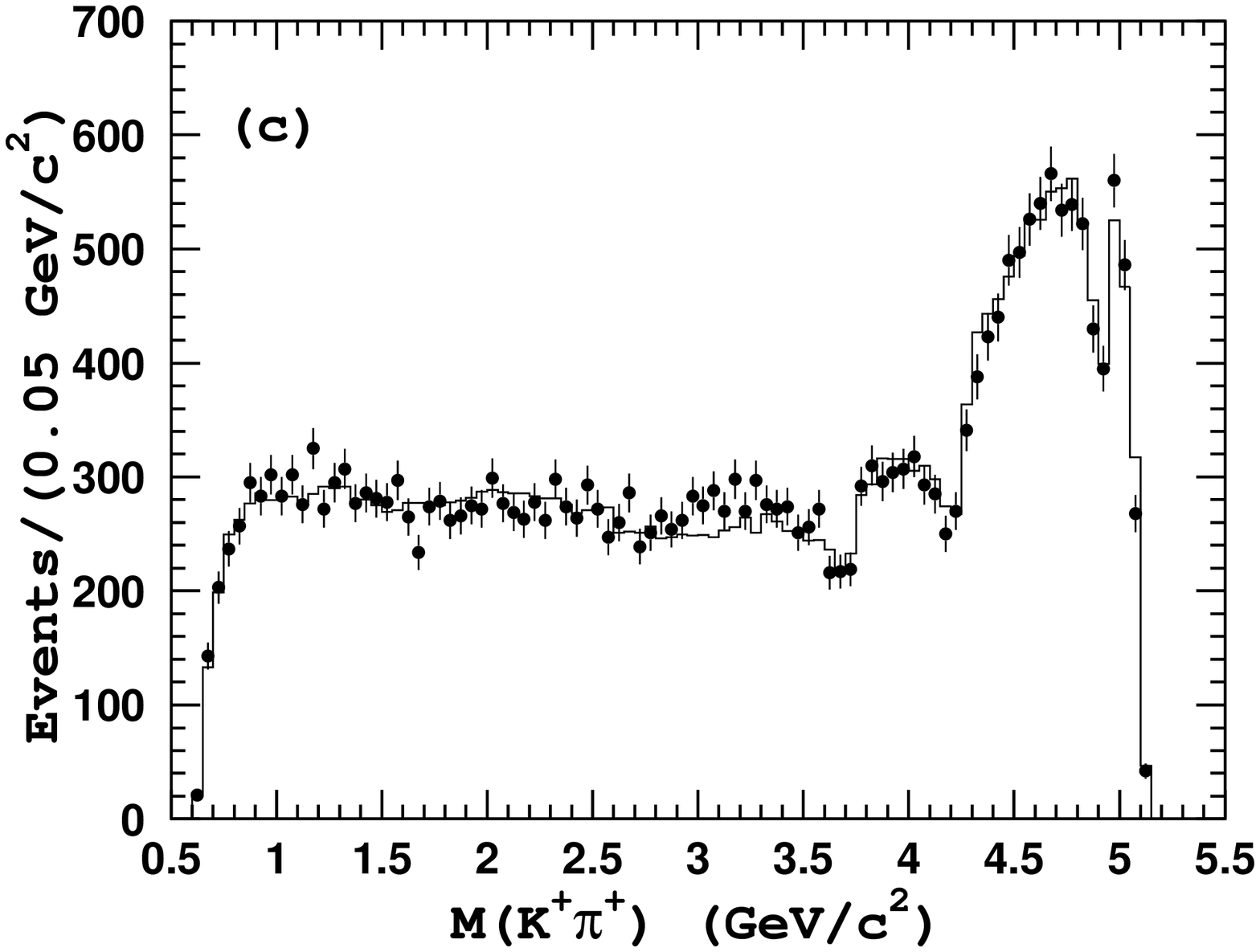}
  \caption{Results of the best fit to the $\kpp$ events in the $\mb-\de$ 
           sidebands shown as projections onto two-particle invariant mass
           variables. Points with error bars are data; histograms
           are fit results.}
\label{fig:khh_back}
\end{figure}

The projections of the data and fits for the background events are shown in
Fig.~\ref{fig:khh_back}. The $\chi^2/N_{\rm bins}$ value of the fit is
$127.6/120$.

\subsection{Fitting the $\bckpp$ Signal}
\label{sec:kpp-sig}

The Dalitz plot for $\kpp$ events in the signal region is shown in 
Fig.~\ref{fig:kpp_dp}(b). There are 7757 events in the signal region that
satisfy all the selection requirements. As found in
Ref.~\cite{khh-dalitz-belle} the $\bpkpp$ decay is well described by a matrix
element that is a coherent sum of $K^*(892)^0\pi^+$, $K^*_0(1430)^0\pi^+$,
$\rho(770)^0K^+$, $f_0(980)K^+$, $f_X(1300)K^+$, $\chic K^+$ quasi-two-body
channels and a non-resonant amplitude. The $f_X(1300)K^+$ channel is added
in order to describe an excess of signal events at $M(\pipi)\simeq 1.3$~\Mass.
With current statistics, the contribution of $f_2(1270)K^+$ is found to be
significant, but not sufficient to fully explain the excess of signal events
in this mass region. In this analysis we modify the model by adding two more
quasi-two-body channels: $f_2(1270)K^+$ and $\omega(782)K^+$ and change the
parameterization of the $f_0(980)$ lineshape from a standard Breit-Wigner
function to a coupled channel Breit-Wigner also known as Flatt\'e
parametrization~\cite{Flatte}. Although the $\omega(782)\to\pipi$ branching
fraction is only $(1.70\pm0.27)$\%~\cite{PDG}, the $\omega(782)$ natural width
is rather narrow. As a result a numerical factor of
$\Gamma(\rho(770)^0)/\Gamma(\omega(782))\sim18$ is introduced in the
$B\to\omega(782)K$ amplitude (relative to the $B\to\rho(770)K$ amplitude) that
compensates the smallness of the $\omega(782)\to\pipi$ branching. As the
independently measured $B^\pm\to\omega(782)K^\pm$ branching
fraction~\cite{HFAG} is comparable to that for $B^\pm\to\rho(770)^0K^\pm$,
the interference between these two amplitudes might significantly distort the
$\rho(770)^0$ lineshape.
\begin{figure}[!t]
  \centering
  \includegraphics[width=0.48\textwidth]{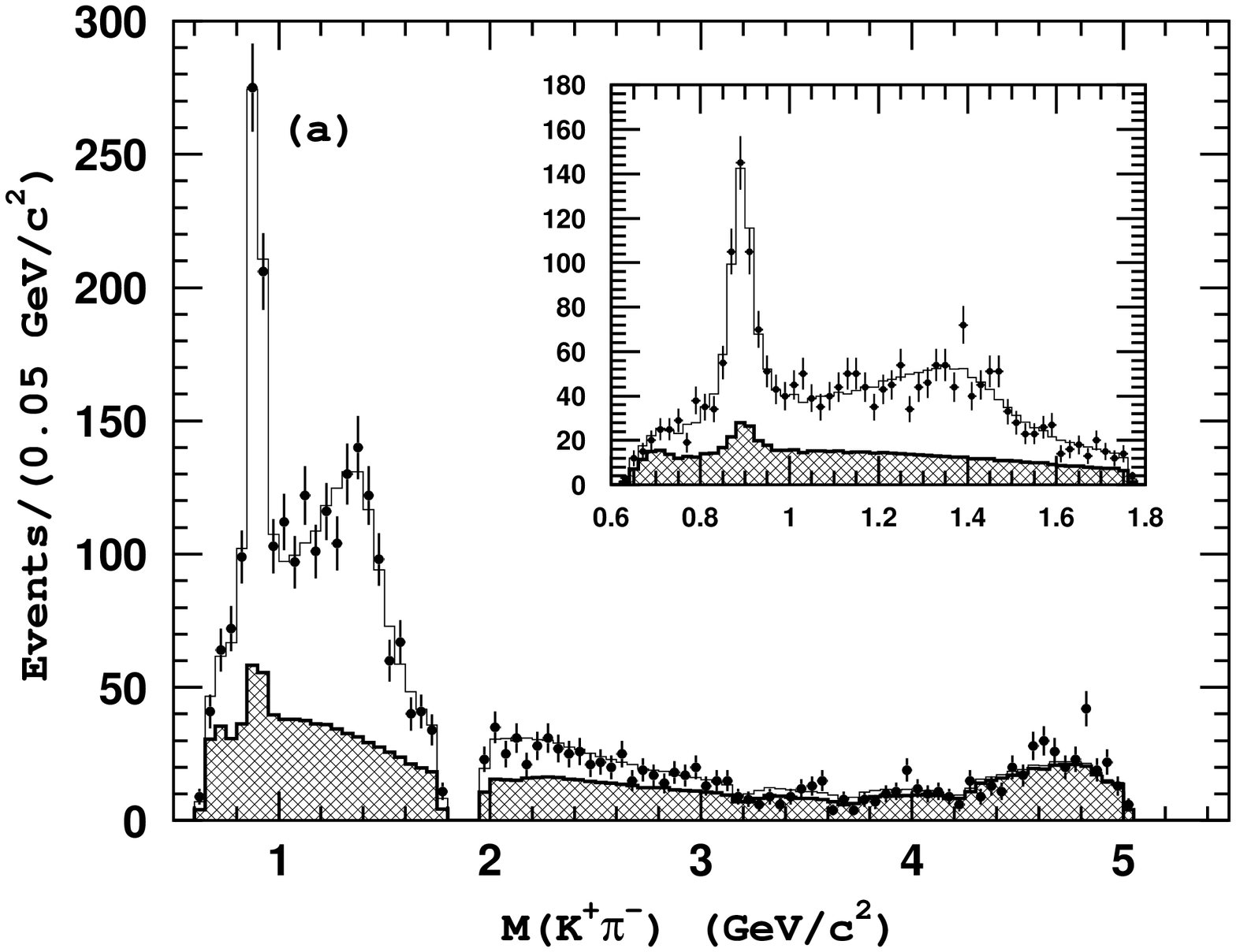} \hfill
  \includegraphics[width=0.48\textwidth]{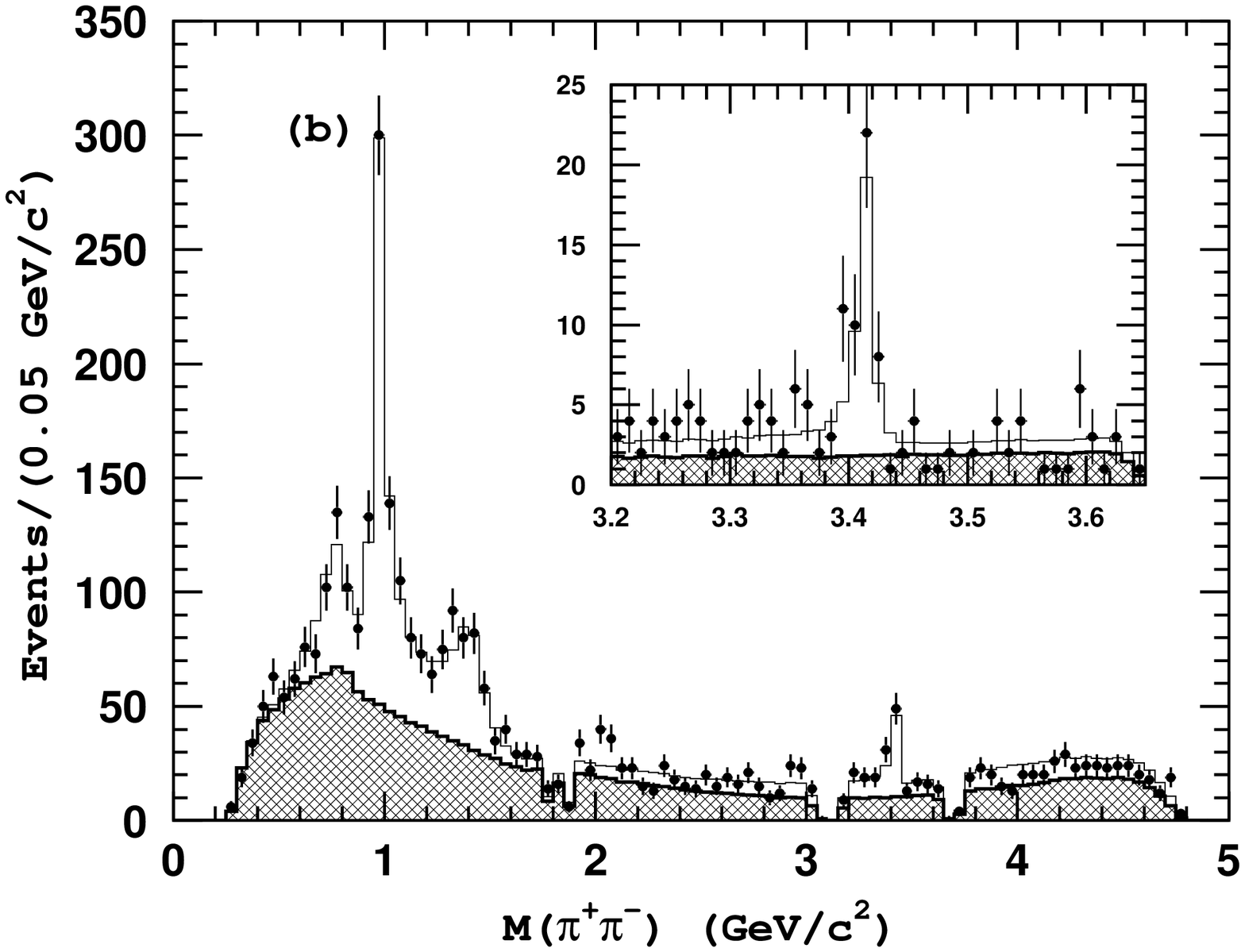}
  \caption{Results of the fit to $\kpp$ events in the signal region.
           Points with error bars are data, the open histograms
           are the fit result and hatched histograms are the background
           components. Inset in (a) shows the $K^*(892)-K_0^*(1430)$
           mass region in 20~\mass~ bins. Inset in (b) shows the
           $\chic$ mass region in 10~\mass~ bins.}
\label{fig:kpp-mod-d0}
\end{figure}
Finally, for $CP$ violation studies the amplitude for each quasi-two-body
channel is modified to include two components: one that is independent of
the sign of the $B$ charge and a component that changes sign with the charge
of the $B$ meson. The resulting decay amplitude reads as
\begin{eqnarray}
&&  {\cal{M}}(B^\pm\to K^\pm\pi^\pm\pi^\mp) = \nonumber \\
&&~~= a_{K^*}e^{i\delta_{K^*}}(1\pm b_{K^*}e^{i\varphi_{K^*}})\Am(K^*(892)^0)
    + a_{K^*_0}e^{i\delta_{K^*_0}}(1\pm b_{K^*_0}e^{i\varphi_{K^*_0}})
                                  \Am(K^*_0(1430)^0) \nonumber \\
&&~~+~ a_{\rho}e^{i\delta_{\rho}}(1\pm b_{\rho}e^{i\varphi_{\rho}})
                                  \Am(\rho(770)^0)
   + a_{\omega}e^{i\delta_{\omega}}(1\pm b_{\omega}e^{i\varphi_{\omega}})
                                  \Am(\omega(782))         \nonumber \\
&&~~+~ a_{f_0}e^{i\delta_{f_0}}(1\pm b_{f_0}e^{i\varphi_{f_0}})
                                  \Am_{\rm Flatte}(f_0(980))
   + a_{f_2}e^{i\delta_{f_2}}(1\pm b_{f_2}e^{i\varphi_{f_2}})
                                  \Am(f_2(1270)) \nonumber \\
&&~~+~ a_{f_X}e^{i\delta_{f_X}}(1\pm b_{f_X}e^{i\varphi_{f_X}})
                                  \Am(f_X)  
   + a_{\chic}e^{i\delta_{\chic}}(1\pm b_{\chic}e^{i\varphi_{\chic}})
                                  \Am(\chic)               \nonumber \\ 
&&~~+~  {\cal A}_{\rm nr}(K^\pm\pi^\pm\pi^\mp)
\label{eq:kpp-amp-d}
\end{eqnarray}
with the non-resonant amplitude ${\cal A}_{\rm nr}$ parametrized by an
empirical function
\begin{equation}
{\cal A}_{\rm nr}(K^\pm\pi^\pm\pi^\mp)
      = a_1^{\rm nr}e^{-\alpha{s_{13}}}e^{i\delta^{\rm nr}_1} 
      + a_2^{\rm nr}e^{-\alpha{s_{23}}}e^{i\delta^{\rm nr}_2},
  \label{eq:kpp-non-res}
\end{equation}
where $s_{13}\equiv M^2(K^\pm\pi^\mp)$, $s_{23}\equiv M^2(\pipi)$. Note that
alternative parameterizations of the non-resonant amplitude
possible~\cite{khh-dalitz-belle,babar-kpp-dcpv}.
The amplitudes $a_i$ and $b_i$, relative phases $\delta_i$ and $\varphi_i$,
mass, $g_{\pi\pi}$ and $g_{KK}$ of the $f_0(980)$, mass and width of the
$f_X(1300)$, and parameter $\alpha$ of the non-resonant amplitude are fit
parameters. With such a parameterization of the amplitude, the $CP$ violating
asymmetry $\ACP$ for a particular quasi-two-body channel can be calculated as
\begin{equation}
\ACP(f)
      = \frac{\BF(B^-\to f^-)-\BF(B^+\to f^+)}{\BF(B^-\to f^-)+\BF(B^+\to f^+)}
      = -\frac{2b\cos\varphi}{1+b^2}.
\label{eq:acp-dcpv}
\end{equation}

\begin{figure}[!t]
\medskip
\medskip
  \centering
  \includegraphics[width=0.49\textwidth]{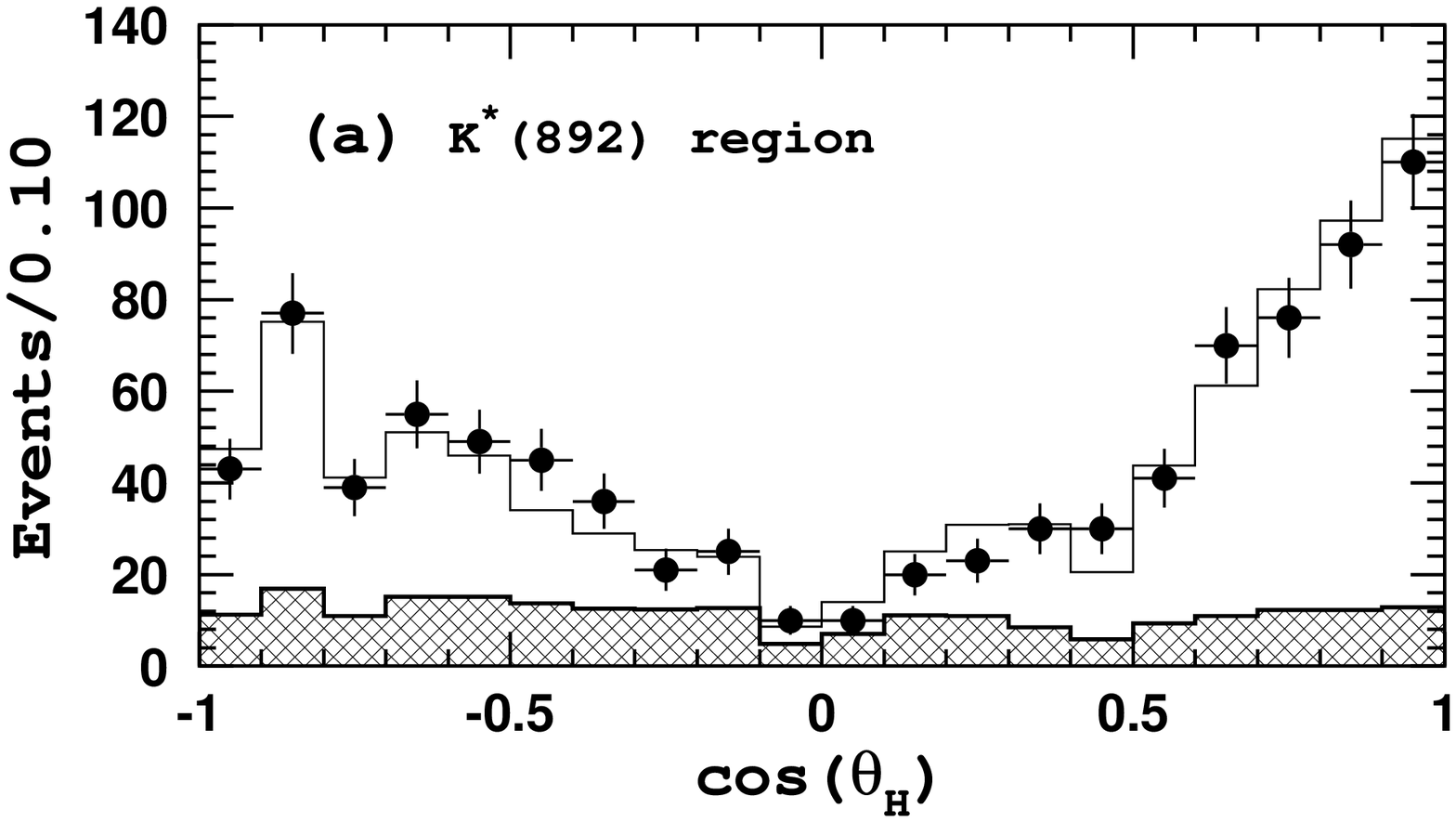} \hfill
  \includegraphics[width=0.49\textwidth]{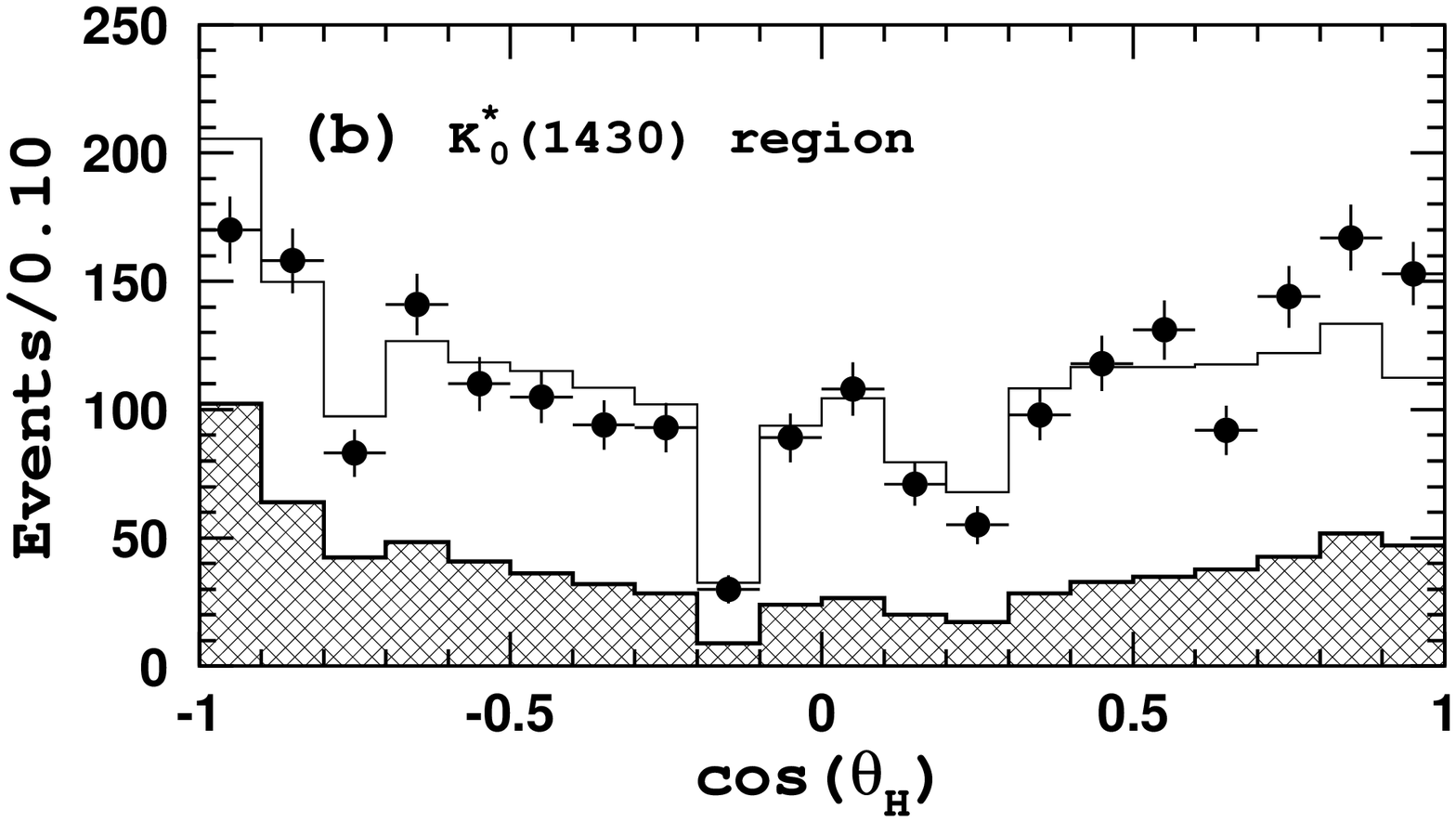} 
  \includegraphics[width=0.49\textwidth]{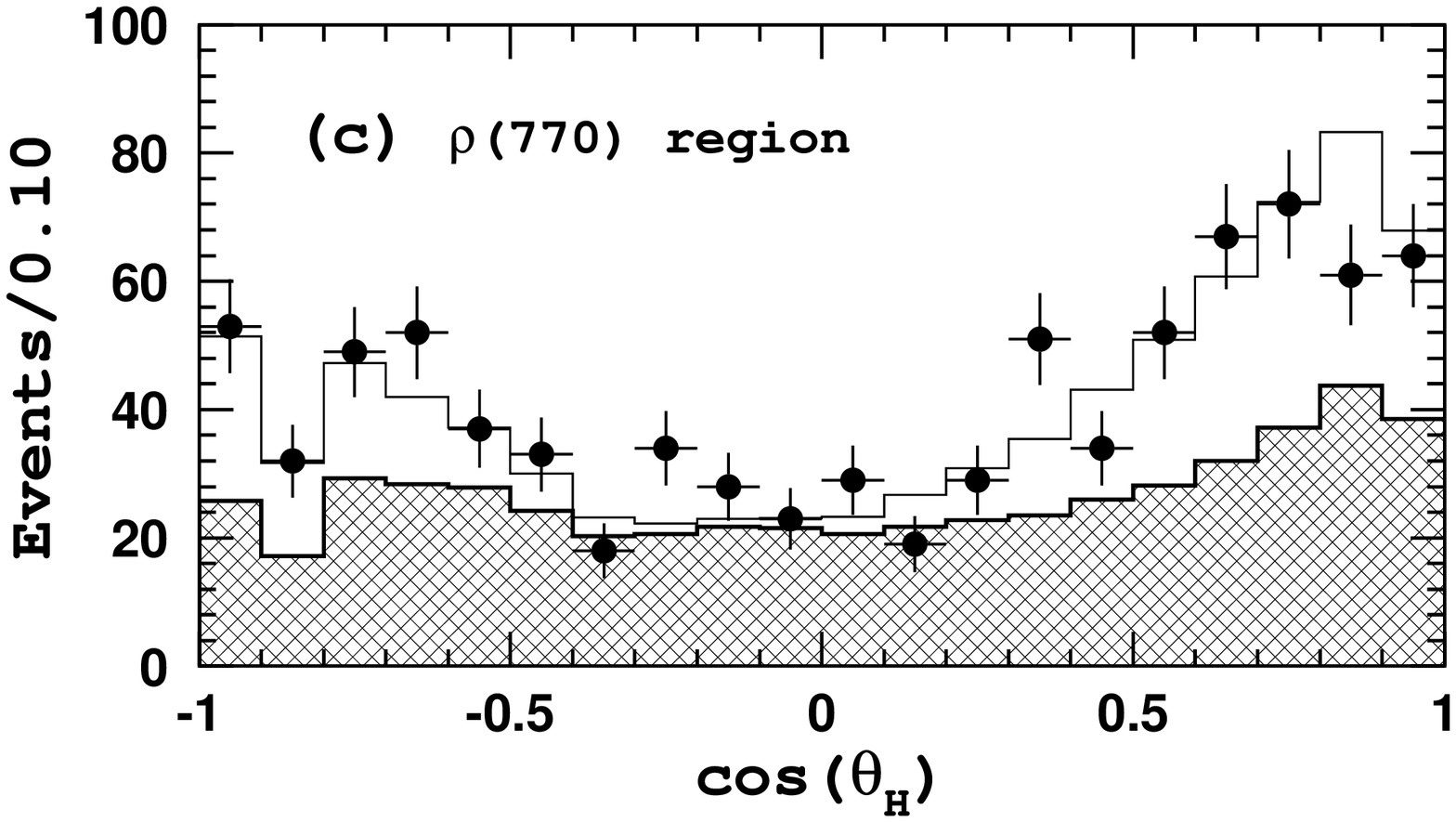} \hfill
  \includegraphics[width=0.49\textwidth]{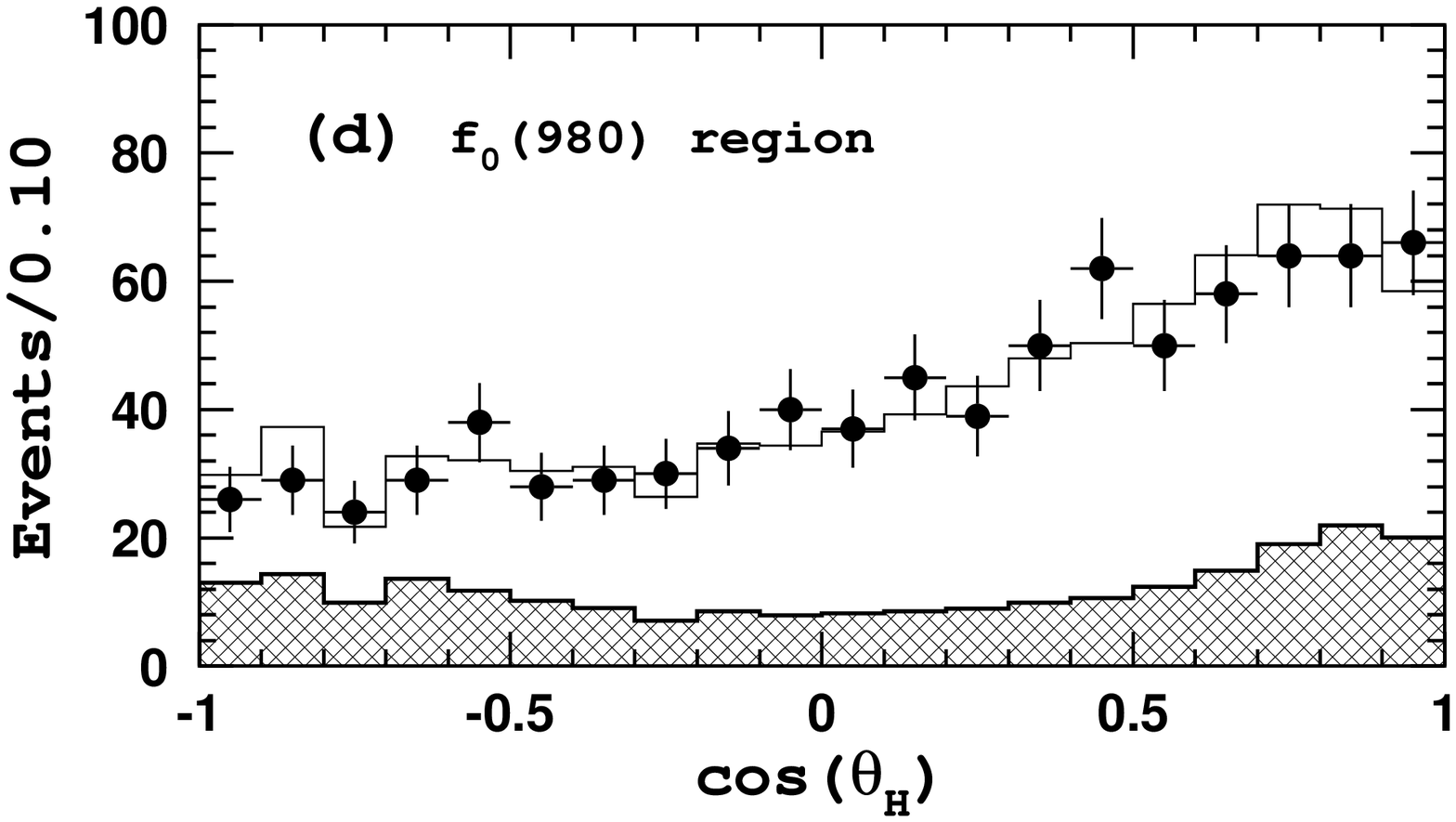}
  \caption{Helicity angle distributions for $\kpp$ events in different regions:
           (a)~$K^*(892)^0$  (0.82~\Mass~$<M(\kcpi)<$~0.97~\Mass);
           (b)~$K^*_0(1430)$ (1.00~\Mass~$<M(\kcpi)<$~1.76~\Mass);
           (c)~$\rho^0(770)$ (0.60~\Mass~$<M(\pipi)<$~0.90~\Mass) and
           (d)~$f_0(980)$    (0.90~\Mass~$<M(\pipi)<$~1.06~\Mass).
           Points with error bars are data, the open histogram is the fit
           result and the hatched histogram is the background component.}
\label{fig:kpp-heli}
\end{figure}

To reduce the number of free fit parameters, we
fit the data in two steps. First we fix all $b_i=0$ and fit the data assuming
no $CP$ violation. From this fit we determine the parameters of the $f_X(1300)$
($M(f_X(1300))=1.449\pm0.013$~\Mass, $\Gamma(f_X(1300))=0.126\pm0.025$~\Mass),
$f_0(980)$ ($M(f_0(980)=0.950\pm0.009$~\Mass, $g_{\pi\pi}=0.23\pm0.05$,
$g_{KK}=0.73\pm0.30$) and the parameter of the non-resonant amplitude 
($\alpha=0.195\pm0.018$). We then fix these parameters and repeat the fit to
data with $b_i$ and $\varphi_i$ floating. In addition, we also assume no $CP$
violation in $B^\pm\to\omega(782) K^\pm$ and for the non-resonant
amplitude. Possible effects of these assumptions are studied and considered
in the final results as a part of the model uncertainty.

\begin{sidewaystable}[!pt]
\caption{Results of the best fit to $\kpp$ events in the $B$ signal region.
The first quoted error is statistical and the second is the model dependent
uncertainty. The quoted significance is statistical only.}
\medskip
\medskip
\label{tab:kpp-fit-res}
\centering
  \begin{tabular}{@{\hspace{2mm}}l@{\hspace{4mm}}c@{\hspace{4mm}}c@{\hspace{4mm}}c@{\hspace{4mm}}c@{\hspace{4mm}}c@{\hspace{4mm}}c@{\hspace{2mm}}}
\hline \hline
Channel & $CP$ averaged  & $\delta$, & $b$ & $\varphi$, & $\ACP$, & Significance,\\
        & fraction, \%   &   degrees  &     & degrees   &  \%  & $\sigma$   \\
\hline \hline
$K^*(892)^0\pi^\pm$  & $13.0\pm0.8^{+0.5}_{-0.7}$
                     & $0$ (fixed)
                     & $0.078\pm0.033^{+0.012}_{-0.003}$
                     & $ -18\pm44^{+5}_{-13}$
                     & $-14.9\pm6.4^{+0.8}_{-0.8}$  & $2.6$ \\
$K_0(1430)^0\pi^\pm$ & $65.5\pm1.5^{+2.2}_{-3.9}$
                     & $55\pm4^{+1}_{-5}$
                     & $0.069\pm0.031^{+0.010}_{-0.008}$
                     & $-123\pm16^{+4}_{-5}$
                     & $+7.5\pm3.8^{+2.0}_{-0.9}$   & $2.7$ \\
$\rho(770)^0K^\pm$   & $ 7.85\pm0.93^{+0.64}_{-0.59}$
                     & $-21\pm14^{+14}_{-19}$
                     & $0.28\pm0.11^{+0.07}_{-0.09}$
                     & $-125\pm32^{+10}_{-85}$
                     & $+30\pm11^{+11}_{-4}$    & $3.9$ \\
$\omega(782)K^\pm$   & $0.15\pm0.12^{+0.03}_{-0.02}$
                     & $100\pm31^{+38}_{-21}$
                     & $0$ (fixed)   & $-$  & $-$  & $-$   \\
$f_0(980)K^\pm$      & $17.7\pm1.6^{+1.1}_{-3.3}$
                     & $67\pm11^{+10}_{-11}$
                     & $0.30\pm0.19^{+0.05}_{-0.10}$
                     & $-82\pm8^{+2}_{-2}$
                     & $-7.7\pm6.5^{+4.1}_{-1.6}$   & $1.6$ \\
$f_2(1270)K^\pm$     & $ 1.52\pm0.35^{+0.22}_{-0.37}$
                     & $140\pm11^{+18}_{-7}$
                     & $0.37\pm0.17^{+0.11}_{-0.03}$
                     & $-24\pm29^{+14}_{-20}$
                     & $-59\pm22^{+3}_{-3}$ & $2.7$ \\
$f_X(1300)K^\pm$     & $ 4.14\pm0.81^{+0.31}_{-0.30}$
                     & $-141\pm10^{+8}_{-9}$
                     & $0.12\pm0.17^{+0.04}_{-0.07}$
                     & $-77\pm56^{+88}_{-43}$
                     & $-5.4\pm16.5^{+10.3}_{-2.4}$  & $1.0$ \\
Non-Res.             & $34.0\pm2.2^{+2.1}_{-1.8}$  
                     & $\delta^{\rm nr}_1=-11\pm5^{+3}_{-3}$
                     & $0$ (fixed)   & $-$ & $-$ & $-$   \\
                     &               
                     & $\delta^{\rm nr}_2=185\pm20^{+62}_{-19}$  \\
$\chic K^\pm$        & $1.12\pm0.12^{+0.24}_{-0.08}$
                     & $-118\pm24^{+37}_{-38}$
                     & $0.15\pm0.35^{+0.08}_{-0.07}$
                     & $-77\pm94^{+154}_{-11}$
                     & $-6.5\pm19.6^{+2.9}_{-1.4}$  & $0.7$ \\
\hline \hline
  \end{tabular}
\end{sidewaystable}

The numerical values of the fit parameters are given in
Table~\ref{tab:kpp-fit-res}. The $\chi^2/N_{\rm bins}$ value of the fit is
$182.5/141$ with $k=32$ fit parameters. Fit projections and the
data are shown in Fig.~\ref{fig:kpp-mod-d0}. Figure~\ref{fig:kpp-heli} shows
helicity angle distributions for several regions, where the helicity angle is
defined as the angle between the direction of flight of the $\pi^-$ in the
$h^+\pi^-$ rest frame and the direction of $B^+$ candidate in the $h^+\pi^-$
rest frame. Gaps visible in Fig.~\ref{fig:kpp-heli} are due to vetoes applied
on invariant masses of two-particle combinations. All plots shown in
Figs.~\ref{fig:kpp-mod-d0} and~\ref{fig:kpp-heli} demonstrate good agreement
between data and the fit.

\begin{figure}[!t]
  \centering
  \includegraphics[width=0.48\textwidth]{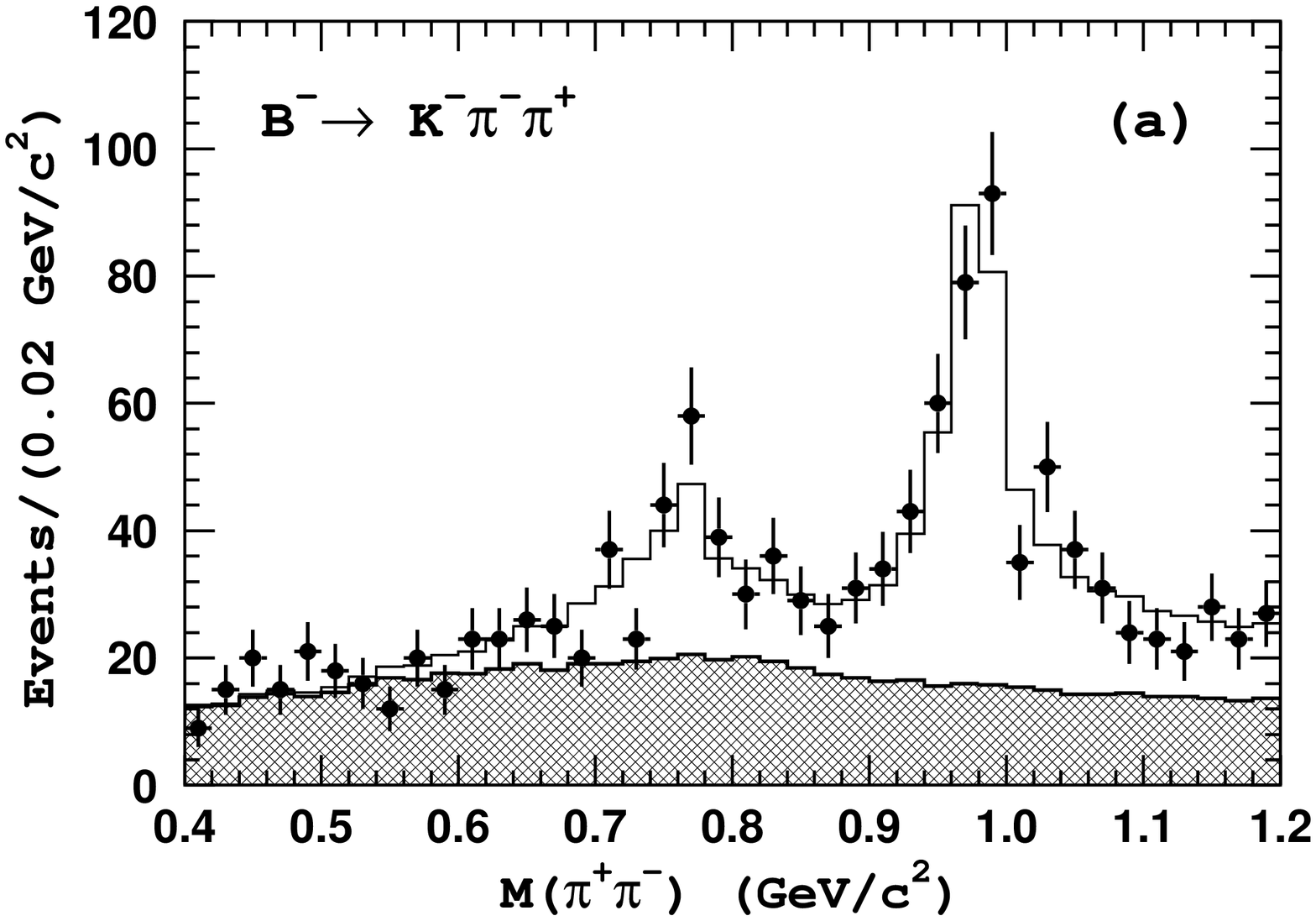} \hfill
  \includegraphics[width=0.48\textwidth]{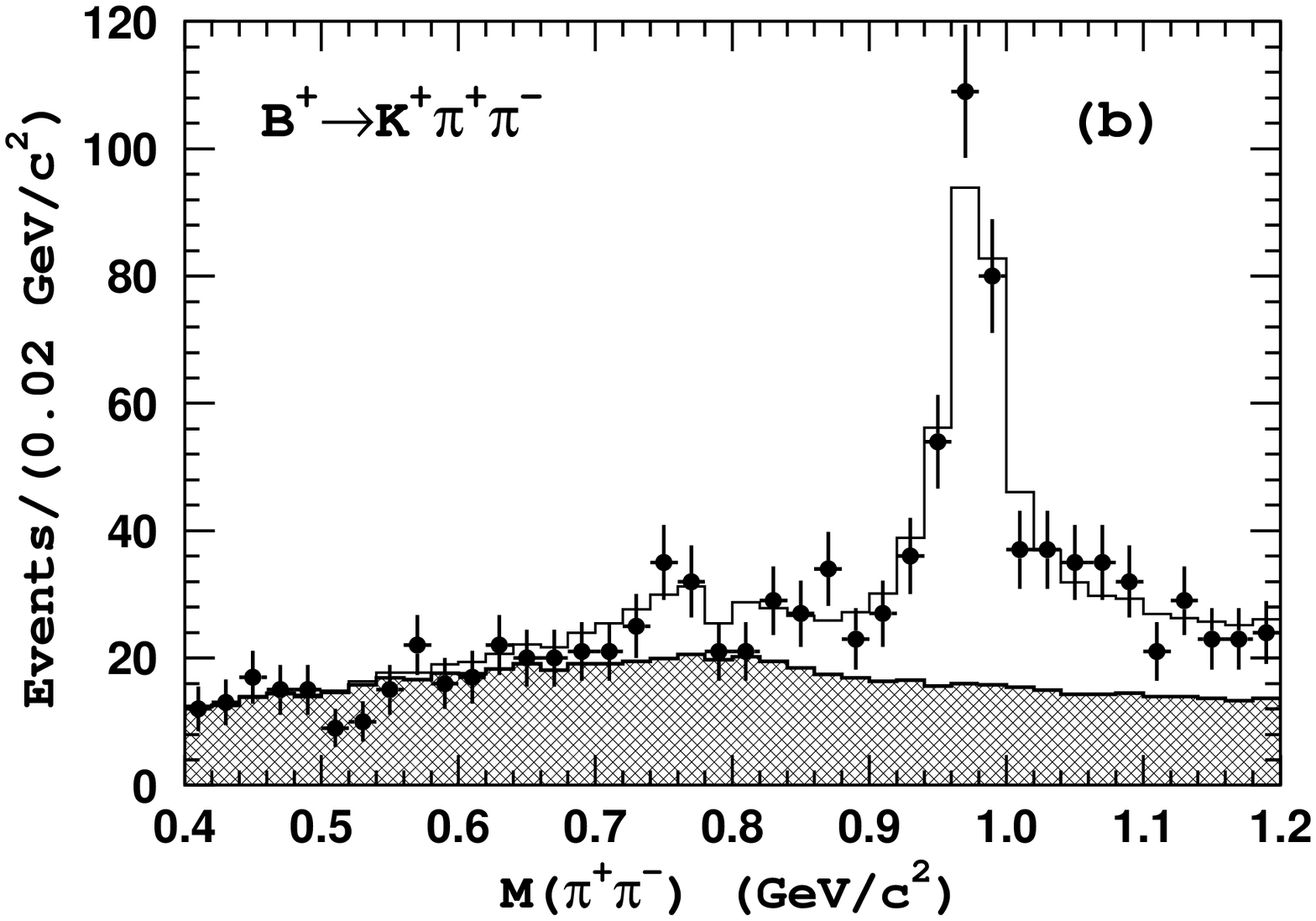} 
  \includegraphics[width=0.48\textwidth]{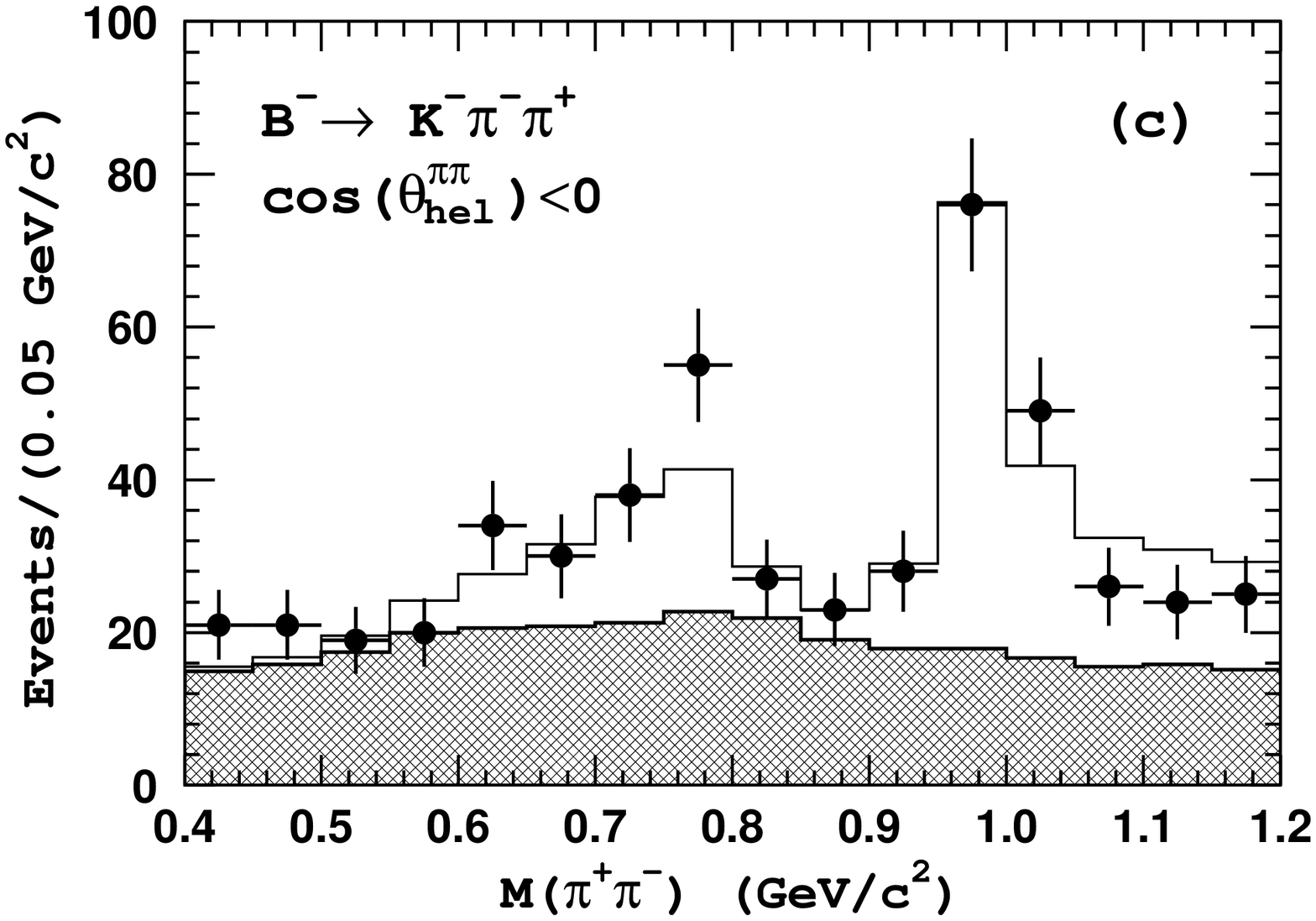} \hfill
  \includegraphics[width=0.48\textwidth]{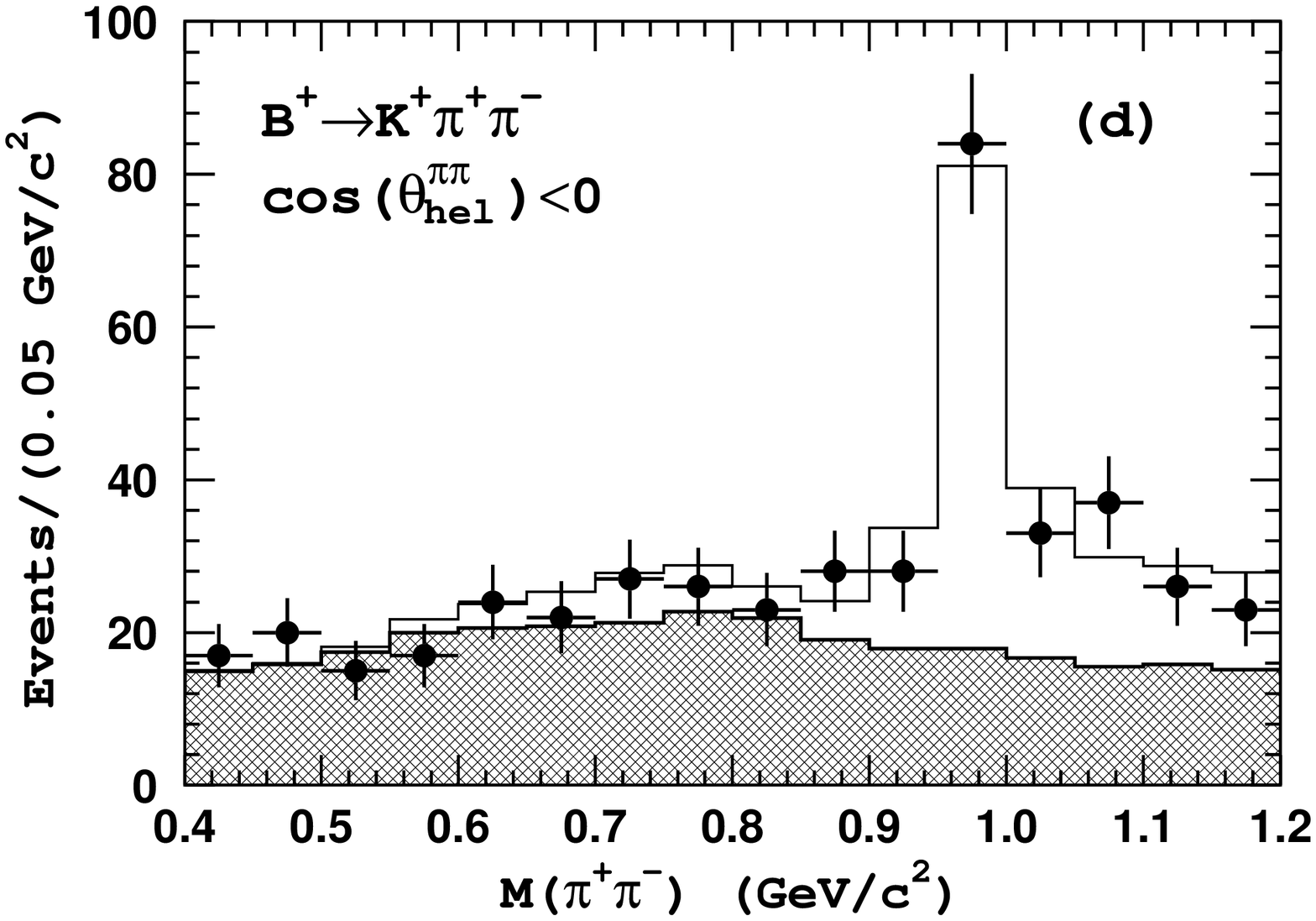}
  \includegraphics[width=0.48\textwidth]{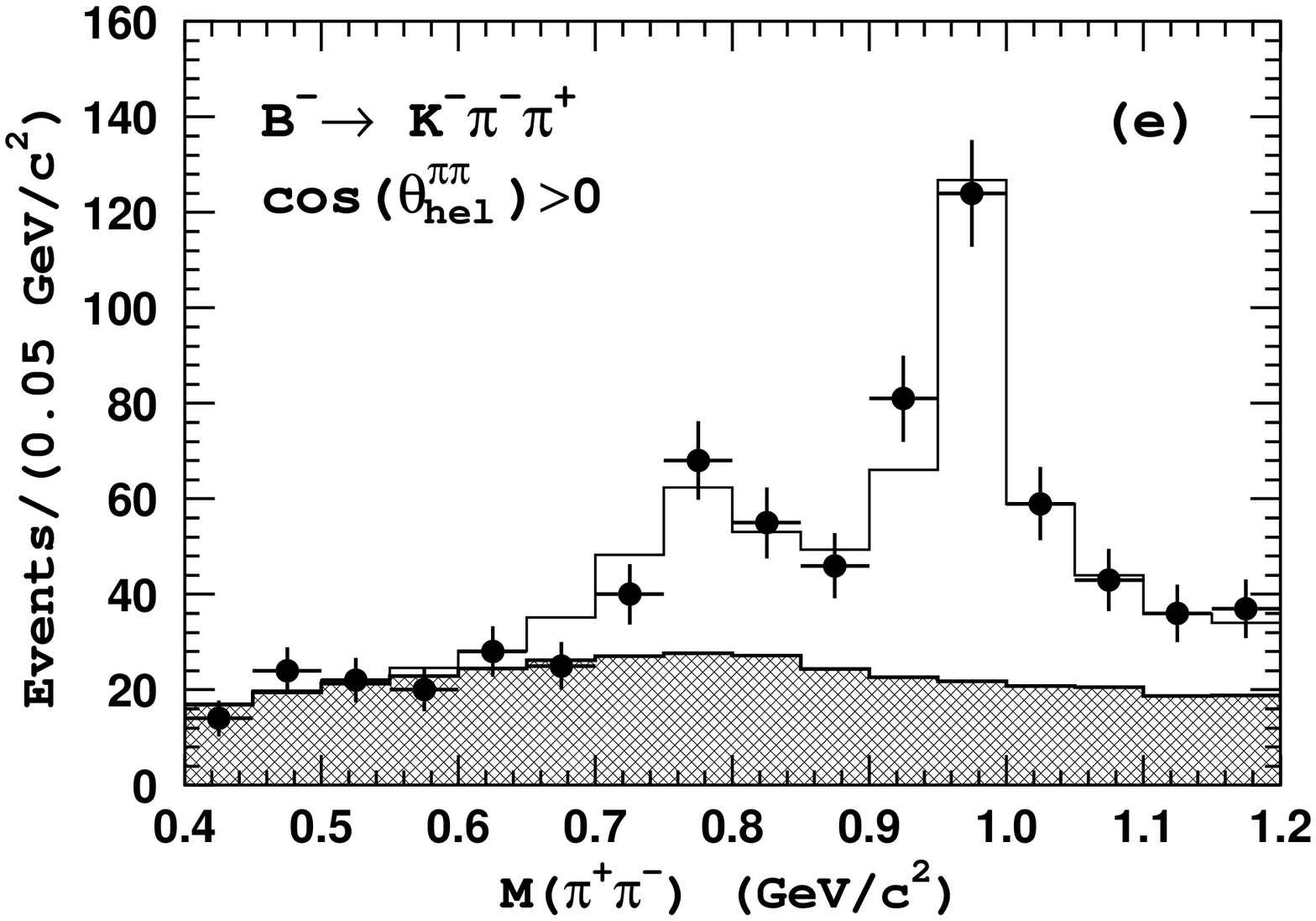} \hfill
  \includegraphics[width=0.48\textwidth]{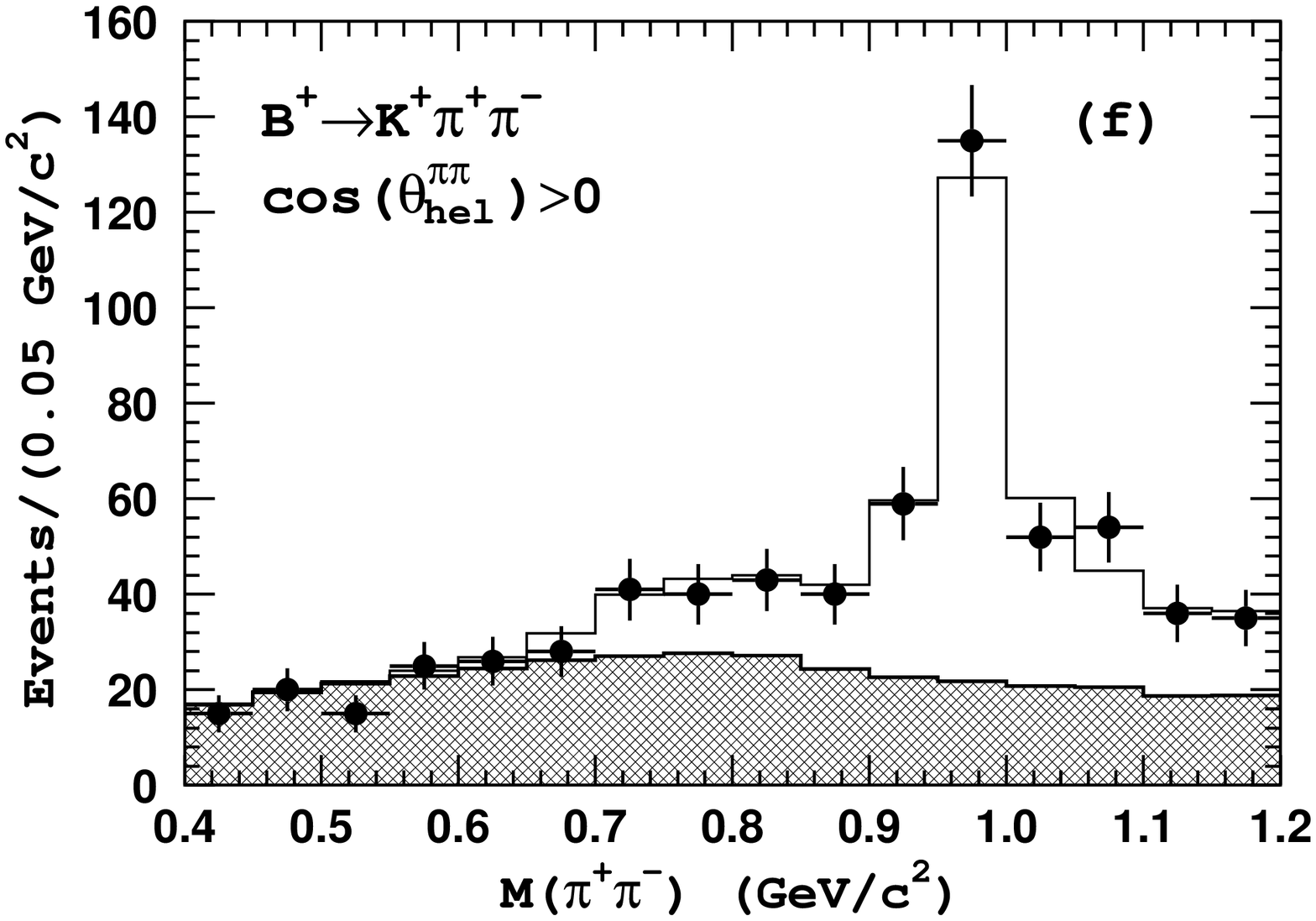}
  \caption{$M(\pipi)$ mass spectra for $B^-$ (left column) and $B^+$ (right
           column) for different helicity regions:
           (a,b) no helicity cuts;
           (c,d) $\cos\theta_H^{\pi\pi}<0$;
           (e,f) $\cos\theta_H^{\pi\pi}>0$;
           Points with error bars are data, the open histogram is the fit
           result and the hatched histogram is the
           background component.}
\label{fig:kpp-pp}
\end{figure}

The statistical significance of the asymmetry quoted in
Table~\ref{tab:kpp-fit-res} is calculated as 
$\sqrt{-2\ln({\cal L}_0/{\cal L}_{\rm max})}$, where ${\cal L}_{\rm max}$ and
${\cal L}_0$ denote the maximum likelihood with nominal fit and with the
asymmetry fixed at zero, respectively. The only channel where the statistical
significance of the $CP$ asymmetry exceeds the $3\sigma$ level is
$B^\pm\to\rho(770)^0K^\pm$. Figures~\ref{fig:kpp-pp}(a,b) show the $\pipi$
invariant mass distributions for the $\rho(770)^0-f_0(980)$ mass region
separately for $B^-$ and $B^+$ events. The effect is even more apparent when
the $M(\pipi)$ distribution for the two helicity angle regions
($\cos\theta_H^{\pi\pi}<0$ and $\cos\theta_H^{\pi\pi}>0$) shown in
Fig.~\ref{fig:kpp-pp}(c-f) are compared.

\section{Systematic \& Model Uncertainties}

The dominant sources of systematic error are listed in Table~\ref{khh_syst}.
The systematic uncertainty in charged track reconstruction is estimated
using partially reconstructed $D^*\to D\pi$ events and from comparison of the
ratio of $\eta\to\pi^+\pi^-\pi^0$ to $\eta\to\gamma\gamma$ events in data and
MC. The uncertainty from the particle identification efficiency is estimated
using pure samples of kaons and pions from $D^0\to K^-\pi^+$ decays, where
the $D^0$ flavor is tagged using $D^{*+}\to D^0\pi^+$ decays. The systematic
uncertainty due to requirements on event shape variables is estimated from a
comparison of the $|\cos\theta_{\rm thr}|$ and ${\cal{F}}$ distributions for
signal MC events and $B^+\to\anti{D}^0\pi^+$ events in the data. We estimate
the uncertainty due to the signal $\de$ shape parameterization by varying the
parameters of the fitting function within their errors. The uncertainty in the
background parameterization is estimated by varying the relative fraction of
the $\bbbar$ background component and the slope of the $\qqbar$ background
function within their errors. 
Reconstruction efficiency is determined using MC events distributed over phase
space according to the matrix element corresponding to the best fit to data.
The relevant systematic uncertainty is estimated to be at the level
of one percent. Finally, to account for variations in reconstruction
efficiency due to modifications in the detector setup and due to non-uniform
data-taking conditions (mainly beam related background conditions), we
generate signal events with background events embedded. Background events are
recorded with random triggers for each experiment. Signal MC events are
generated for each experiment with statistics proportional to experimental
data. The overall systematic uncertainty for the three-body branching fraction
is $7.4\%$.

\begin{table}[!t]
\centering
\caption{List of systematic errors (in percent) for the
         three-body $\bckpp$ branching fraction.}
\medskip
\label{khh_syst}
  \begin{tabular}{lc}  \hline \hline
  Source~\hspace*{92mm} &  \multicolumn{1}{c}{Error}  \\
\hline 
 Charged track reconstruction &     $3.0$          \\
 PID                          &     $4.5$          \\
 Event Shape requirements     &     $2.5$          \\
 Signal yield  extraction     &     $3.9$          \\
 Model                        &     $1.2$          \\
 MC statistics                &     $1.0$          \\
 Luminosity measurement       &     $1.0$          \\
\hline
 Total                        &     $7.4$          \\
\hline \hline
  \end{tabular}
\end{table}

\begin{table}[!b]
\centering
\caption{List of systematic errors for CP
         violating asymmetry.}
\medskip
\label{tab:acp_syst}
  \begin{tabular}{lc}  \hline \hline
  Source~\hspace*{92mm} &  \multicolumn{1}{c}{$\delta \ACP$} \\
\hline 
 Rare background              &                      \\
 ~~~~$\rho^0\pi^\pm$          &     $+0.003/-0.004$  \\
 ~~~~$\eta'K^\pm$             &     $+0.001/-0.001$  \\  
 ~~~~$K^\pm\pi^\mp$           &     $+0.004/-0.004$  \\
 Detector asymmetry           &     $+0.023/-0.023$  \\
 Signal yield  extraction     &     $+0.011/-0.011$  \\
\hline
 Total                        &     $+0.031/-0.029$  \\
\hline \hline
  \end{tabular}
\end{table}

Note that in asymmetry calculation most of these systematic uncertainties
cancel. The few remaining sources are listed in Table~\ref{tab:acp_syst}.
Systematic uncertainty due to possible asymmetry in background from charmless
$B$ decays is estimated by introducing an asymmetry equal to experimentally
measured central value~\cite{HFAG} increased by one standard deviation 
to each charmless background component one by one and refitting the data.
The possible bias due to intrinsic detector asymmetry in reconstruction of
tracks of different charges is estimated using $B\to D\pi$ events in data.

To estimate the model dependent uncertainty in the branching fractions and
asymmetries for individual quasi-two-body intermediate states, we vary the
default model and repeat the fit to data. Namely, we add one additional
quasi-two-body channel which is either $K^*(1410)^0\pi^+$, $K^*(1680)^0\pi^+$,
or $K^*_2(1430)^0\pi^+$ or remove $\omega(782)K^+$ or $f_2(1270)K^+$ channel
from the default model, use different assumptions on the spin of the
$f_X(1300)$ state and use different parameterizations of the non-resonant
amplitude. For estimation of the model uncertainty in charge asymmetries for
individual quasi-two-body channels in addition to model variations we fit the
data with different assumptions on $CP$ violation in different channels.
Finally, we check the consistency of the $\ACP$ results with those obtained
from independent fits of $B^-$ and $B^+$ subsamples.


\section{Branching Fraction \& Charge Asymmetry Results}
\label{sec:dcpv}

   In the preceding section we determined the relative fractions of various
quasi-two-body intermediate states in the three-body $\bckpp$ decay. To
translate those numbers into absolute branching fractions, we first need to
determine the branching fraction for the three body decay. To determine the
reconstruction efficiency, we use MC simulation where events are distributed
over the phase space according to the matrix elements obtained from the best
fit to data. The corresponding reconstruction efficiency is
$(22.4\pm0.2)$\%. Results of the branching fraction and $CP$-violating
asymmetry calculations are summarized in Table~\ref{tab:results}.

\begin{table}[!ht]
  \caption{Summary of branching fraction results. The first quoted error is
           statistical, the second is systematic and the third is the model
           uncertainty.}
  \medskip
  \label{tab:results}
\centering
  \begin{tabular}{lccr} \hline \hline
Mode & {\small $\BF(B^+\to Rh^+\to \kppp)$} &
$\BF(B^+\to Rh^+)$ & \multicolumn{1}{c}{$A_{CP},$} \\
 & $\times10^{6}$ & $\times10^{6}$ & \multicolumn{1}{c}{\%}
 \\ \hline \hline
 $\kpp$ Charmless       &    $-$
                        & $48.8\pm1.1\pm3.6$
                        & $4.9\pm2.6\pm3.0$  \\
 $K^*(892)^0[K^+\pi^-]\pi^+$ 
                        & $6.45\pm0.43\pm0.48^{+0.25}_{-0.35}$
                        & $9.67\pm0.64\pm0.72^{+0.37}_{-0.52}$ 
                        & $-14.9\pm6.4\pm3.0^{+0.8}_{-0.8}$  \\
 $K^*_0(1430)[K^+\pi^-]\pi^+$
                        & $32.0\pm1.0\pm2.4^{+1.1}_{-1.9}$
                        & $51.6\pm1.7\pm6.8^{+1.8}_{-3.1}$
                        & $+7.6\pm3.8\pm3.0^{+2.0}_{-0.9}$   \\
 $\rho(770)^0[\pi^+\pi^-]K^+$
                        & $3.89\pm0.47\pm0.29^{+0.32}_{-0.29}$
                        & $3.89\pm0.47\pm0.29^{+0.32}_{-0.29}$
                        & $+30\pm11\pm3.0^{+11}_{-4}$     \\
 $f_0(980)[\pi^+\pi^-]K^+$
                        & $8.78\pm0.82\pm0.65^{+0.55}_{-1.64}$
                        & $-$                                 
                        & $-7.7\pm6.5\pm3.0^{+4.1}_{-1.6}$   \\
 $f_2(1270)[\pi^+\pi^-]K^+$
                        & $0.75\pm0.17\pm0.06^{+0.11}_{-0.18}$
                        & $1.78\pm0.41\pm0.14^{+0.26}_{-0.43}$
                        & $-59\pm22\pm3.0^{+3}_{-3}$   \\
 Non-resonant
                        & $-$
                        & $16.9\pm1.3\pm1.3^{+1.1}_{-0.9}$
                        & $-$     \\
\hline
 $\chic[\pi^+\pi^-]K^+$
                        & $0.56\pm0.06\pm0.04^{+0.12}_{-0.04}$
                        & $112\pm12\pm18^{+24}_{-8}$
                        & $-6.5\pm19.6\pm3.0^{+2.9}_{-1.4}$     \\
\hline \hline
  \end{tabular}
\end{table}


\section{Discussion \& Conclusion}
\label{sec:discus}

  With a 357~fb$^{-1}$ data sample collected with the Belle detector, we made
the first analysis of direct $CP$ violation in the three-body charmless decay
$\bckpp$. Results on branching fraction and $CP$-violating asymmetry
calculations are summarized in Table~\ref{tab:results}. In all except the
$B^\pm\to\rho(770)^0K^\pm$ channel the measured asymmetry is below $3\sigma$
statistical significance. Evidence for large direct $CP$ violation the decay 
$B^\pm\to\rho(770)^0K^\pm$ is found in agreement with our results obtained
with 253~fb$^{-1}$~\cite{belle-kpp-dcpv} and with results by
BaBar~\cite{babar-kpp-dcpv}. This is also in agreement with some theoretical
predictions~\cite{beneke-neubert}. The statistical significance of the
asymmetry observed in $B^\pm\to\rho(770)^0K^\pm$ is $3.9\sigma$. Depending on
the model used to fit the data the significance varies from $3.7\sigma$ to
$4.0\sigma$. If confirmed with a larger data sample, this would be the first
observation of $CP$ violation in the decay of a charged meson.


\section*{Acknowledgments}

We thank the KEKB group for the excellent operation of the
accelerator, the KEK cryogenics group for the efficient
operation of the solenoid, and the KEK computer group and
the National Institute of Informatics for valuable computing
and Super-SINET network support. We acknowledge support from
the Ministry of Education, Culture, Sports, Science, and
Technology of Japan and the Japan Society for the Promotion
of Science; the Australian Research Council and the
Australian Department of Education, Science and Training;
the National Science Foundation of China under contract
No.~10175071; the Department of Science and Technology of
India; the BK21 program of the Ministry of Education of
Korea and the CHEP SRC program of the Korea Science and
Engineering Foundation; the Polish State Committee for
Scientific Research under contract No.~2P03B 01324; the
Ministry of Science and Technology of the Russian
Federation; the Ministry of Higher Education, 
Science and Technology of the Republic of Slovenia;  
the Swiss National Science Foundation; the National Science Council and
the Ministry of Education of Taiwan;
and the U.S.\ Department of Energy.


\end{document}